\documentclass{LMCS}

\usepackage{graphicx,color}
\usepackage{stmaryrd}
\usepackage{amsfonts}

\usepackage{mathtools}
\usepackage{enumerate}
\usepackage{hyperref}

\input xy
\xyoption{all}

\itemsep\smallskipamount\parsep0pt\topsep\medskipamount
\newcounter{lister}

{\end{list}}
\newcommand{\ealist}{\end{gqenumerate}\noindent}

\itemsep\smallskipamount\parsep0pt\topsep\medskipamount
\newcounter{lister2}

{\end{list}}
\newcommand{\ealistb}{\end{gqitemize}\noindent}

\usepackage{color}

\let\pf\proof
\let\epf\endproof

\newtheorem{ex}[thm]{Example}
\newtheorem{const}[thm]{Construction}

\newtheorem{defn}{Definition}[section]

\newcommand{\chum}[3]{\xymatrix @=6mm{ \ar@<0.4ex>[r]^{{#1}_{#3}} & \ar@<0.4ex>[l]^{{#2}_{#3}}}}
\newcommand{\cphi}[2]{\xymatrix @=6mm{ \ar@<0ex>[r]^{{\smash{#1}}_{\smash{#2}}} &}}

\def\doi{6 (1:3) 2010}
\lmcsheading%
{\doi}
{1--20}
{}
{}
{Apr.~16, 2007}
{Jan.~14, 2010}
{}   

\begin{document}

\title{Bifinite Chu Spaces}

\author[M.~Droste]{Manfred Droste}
\address{Institute of Computer Science,
Leipzig University,
04158 Leipzig, Germany}
\email{droste@informatik.uni-leipzig.de}

\author{Guo-Qiang Zhang}\
\address{Department of Electrical Engineering and Computer Science,
Case Western Reserve University,
Cleveland, OH 44106, U.S.A.}
\email{gq@case.edu}

\begin{abstract}
   This paper studies colimits of sequences of finite Chu spaces and
   their ramifications.  Besides generic Chu spaces, we consider
   extensional and biextensional variants.  In the corresponding
   categories we first characterize the monics and then the existence
   (or the lack thereof) of the desired colimits.  In each case, we
   provide a characterization of the finite objects in terms of
   monomorphisms/injections.  Bifinite Chu spaces are then expressed
   with respect to the monics of generic Chu spaces, and universal,
   homogeneous Chu spaces are shown to exist in this category.
   Unanticipated results driving this development include the fact that
   while for generic Chu spaces monics consist of an injective first
   and a surjective second component, in the extensional and
   biextensional cases the surjectivity requirement can be dropped.
   Furthermore, the desired colimits are only guaranteed to exist in
   the extensional case.  Finally, not all finite Chu spaces
   (considered set-theoretically) are finite objects in their
   categories.  This study opens up opportunities for further
   investigations into recursively defined Chu spaces, as well as
   constructive models of linear logic.
\end{abstract}



\maketitle

\section{Introduction}\label{intro}

\noindent Within semantic frameworks for programming languages, a
basic approach to the study of infinite objects is through their
finite approximations. This is true both within an individual domain,
as well as with domains collectively. A salient example of the latter
is Plotkin's approach to SFP~\cite{plotkin}, where a class of domains
is constructed systematically by taking colimits of sequences of
finite partial orders.  An important component of this framework is
the notion of embedding-projection pair, capturing when one partial
order is an approximation of another.  An interesting outcome of this
process is that completeness of an individual domain, the property
that makes a cpo complete, becomes a natural by-product obtained by
taking colimits of finite structures.  In domain theory, the SFP- (or
bifinite) domains now form an important cartesian closed category of
domains, see~\cite{curien}.

In this paper we study colimits of sequences of finite Chu spaces.
This entails the use of monic morphisms (or monomorphisms) as a way to
formulate the substructure relationship.  Such an innocuous attempt
led to striking differentiations of the notion dictated by the
extensionality properties of the underlying spaces. We consider three
base categories of Chu spaces: the generic Chu spaces ({\bf C}), the
extensional Chu spaces ({\bf E}), and the biextensional Chu spaces
({\bf B}).  The main results are: (1) a characterization of monics in
each of the three categories; (2) existence (or the lack thereof) of
colimits and a characterization of finite objects in each of the
corresponding categories using monomorphisms/injections (denoted as
{\bf iC}, {\bf iE}, and {\bf iB}, respectively); (3) a formulation of
bifinite Chu spaces with respect to {\bf iC}; (4) the existence of
universal, homogeneous Chu spaces in this category.  Unanticipated
results driving this development include the fact that: (a) in {\bf
  C}, a morphism $(f,g)$ is monic iff $f$ is injective and $g$ is
surjective while for {\bf E} and {\bf B}, $(f,g)$ is monic iff $f$ is
injective (but $g$ is not necessarily surjective); (b) while colimits
always exist in {\bf iE}, it is not the case for {\bf iC} and {\bf
  iB}; (c) not all finite Chu spaces (considered set-theoretically)
are finite objects in their categories.

Bifinite Chu spaces can be viewed, in an intuitive category-theoretic
sense, as ``countable'' objects which are approximable by the finite
objects of the category. The class of bifinite Chu spaces is very rich
(up to isomorphism, there are uncountably many such spaces). However,
we show that there is a single bifinite Chu space $U$ which contains
any other bifinite Chu space as a subspace. Moreover, $U$ can be
chosen to be homogeneous, i.e. to bear maximal possible degree of
symmetry, and with this additional property $U$ is unique up to
isomorphism.

Our interest in Chu spaces stems from a number of recent developments.
Chu spaces provide a suitable model of linear logic, originating from
a general categorical construction introduced by Barr and his
student~\cite{barr1,barr2}.  The rich mathematical content of Chu
spaces has been extensively illustrated by Pratt and his collaborators
in a variety of settings, ranging from concurrency to logic and
category theory~\cite{pratt1,pratt2,pratt3,pratt4,pratt6,pratt7}.  In
particular, Pratt shows that all small categories can be embedded in
a certain category of Chu spaces~\cite{pratt5}.

Chu spaces are closely related to the topic of Formal Concept
Analysis (FCA~\cite{ganter,zhang-mfps}).  Both areas use the same
objects but the morphisms considered in FCA are different.
Chu spaces are also related to domains~\cite{zhang-mfps}.  In
\cite{lamarche}, a class called casuistries was introduced as a
``continuous'' version of Chu spaces, and yet maintaining the
constructions desired as a model of linear logic. On the other hand,
if instead of Chu transformations, Chu spaces are equipped with what
are called {\em approximable mappings}~\cite{pascal,scott}, one
obtains a cartesian closed category equivalent to the category of
algebraic lattices and Scott continuous functions~\cite{alga}. For
this to work properly, a modified notion of formal concept, called
{\em approximable concept}, needs to be used~\cite{tac}. This way, an
infinite concept can be approximated by finite ones.

Universal objects have played an important role in the development of
domain theory. For example,
the early work of Scott~\cite{scott76} and Plotkin~\cite{plotkin78}
showed that with universal objects, domain equations can be
treated by a calculus of retracts.

By studying Chu spaces that  are colimits of sequences of finite objects,
we hope to understand these spaces from a constructive angle, formulate a notion of
completeness, and study the existence of universal, homogeneous objects.
Recursively defined Chu spaces as well as models of linear logic within bifinite Chu
spaces are some topics worth revisiting in light of this paper.

\smallskip

The rest of the paper is organized as follows. Section 2 recalls basic terminologies and
gives a  characterization of monic morphisms in the categories {\bf C}, {\bf E}, and {\bf B}.
Section 3 studies colimits in the categories  {\bf iC}, {\bf iE}, and {\bf iB}.
Section 4 characterizes finite objects in  {\bf iC}, {\bf iE}, and {\bf iB}.
Section 5 introduces bifinite Chu spaces
and shows the existence of universal, homogeneous bifinite Chu spaces using
finite amalgamation.  

Remark: The shortened conference version of the paper was presented at CALCO 2007 in Bergen, Norway.

\section{Chu spaces and monic morphisms}

\noindent We recall some basic definitions to fix notation,
following~\cite{pratt0}.  Readers interested in more details should
consult~\cite{pratt2}.  In~\cite{pratt2}, the morphisms are called Chu
transformations.

\begin{defn}
A Chu space over a set $\Sigma$ is a triple $(A, r, X)$
where $A$ is a set whose elements can be considered as {\em objects} and $X$ is a set
whose elements can be regarded as {\em attributes}.
The satisfaction relation $r$ is a
function $A \times X\to \Sigma$.
A morphism   from a Chu space $(A, r, X)$
to a Chu space $(B, s, Y)$
is a pair of functions $(f, g)$, with
$f: A\rightarrow B$ and $g: Y\rightarrow X$
such that for any $a\in A$ and $y\in Y$,
$s(f(a),y) = r(a, g(y)).$
To alleviate the notational burden, we refer to a morphism
by $\varphi = (f,g)$, and refer to the forward component by
$\varphi^+ =f$ and the backward component by $\varphi^-=g$.
\end{defn}

For all the examples we consider in this paper, $\Sigma = \{0,1\}$.
If $\Sigma$ is left unspecified, then it is assumed to contain at least two
elements, denoted as $0$ and $1$.
A Chu space  $(A, r, X)$ has two equivalence relations built-in.
One is on the rows, where the $a$-th row corresponds
to a function $r(a, -) : X\to \Sigma$.
Two rows $a,b$ are {\em equivalent} if $r(a, -)=r(b, -)$. Similarly,
an equivalence relation exists on columns, defined by equality $r(-, x)=r(-, y)$
for $x, y \in X$.
A Chu space  $(A, r, X)$ is  called {\em extensional} if $r(-, x) =r(-, y)$ implies $x=y$,
i.e., $r$ does not contain repeated columns. Similarly, a  Chu space $(A, r, X)$
is {\em separable} if it does not contain repeated rows. Using  topological
analogy, if we think of objects in $A$ as points and attributes in $X$ as open sets,
then separable Chu spaces are those for which distinct points can be differentiated by
the open sets containing them (such spaces are called $T_0$).
A Chu space is {\em biextensional} if it is
both separable and extensional.

We denote by {\bf C}  the category of Chu spaces and morphisms defined above, and {\bf E} and
${\bf B}$ the full subcategories of extensional and biextensional  Chu spaces, respectively.
Composition of morphisms reduces to functional compositions of the components:
$\varphi_2\circ \varphi_1 = (\varphi^+_2\circ \varphi^+_1, {\varphi_1}^{-} \circ  {\varphi_2}^{-} )$,
noting that the second component goes backwards.
For abbreviation, objects are denoted as $\mathsf{C}_i$ for short, where $\mathsf{C}_i := (A_i, r_i, X_i)$.
We refer to $A_i$ the {\em object set}, and $X_i$ as the {\em attribute set} of $\mathsf{C}_i$, respectively.
As a refinement of an observation in~\cite{lamarche}, we have the following
result which will be useful for subsequent developments of the paper.

\begin{prop}\label{lamarche}
Suppose $\varphi_1, \varphi_2: \mathsf{C}\to  \mathsf{C}'$ are morphisms in {\bf C}.
Then
\begin{enumerate}[\em(1)]
\item if   $\mathsf{C}$ is extensional, then $\varphi_1^{+} = \varphi_2^{+} $ implies $\varphi_1^{-} = \varphi_2^{-}$;
\item if   $\mathsf{C}'$  is separable, then $\varphi_1^{-} = \varphi_2^{-} $ implies $\varphi_1^{+} = \varphi_2^{+}$;
\item if   $\mathsf{C}$ and $\mathsf{C}'$ are biextensional, then $\varphi_1^{-} = \varphi_2^{-} $ iff $\varphi_1^{+} = \varphi_2^{+}$.
\end{enumerate}
\end{prop}

Thus,  the forward and backward components in a morphism determine each other
uniquely in the category of biextensional Chu spaces.

\pf
Let us write $\mathsf{C}=(A, r, X)$ and $\mathsf{C}'=(A', r', X')$.  First
we show (1). Suppose $\mathsf{C}$ is extensional and $ \varphi_1^{+} =
\varphi_2^{+}$. Then for all $x' \in X'$ and $a \in A$ we have
$$
   r(a, \varphi_1^{-} (x'))= r'(\varphi_1^{+}(a), x') = r'(\varphi_2^{+}(a),x') = r(a,\varphi_2^{-} (x'))
$$
Hence $\varphi_1^{-} (x') = \varphi_2^{-} (x')$ by extensionality of
$\mathsf{C}$. Now (2) follows from (1) by duality, and (1) and (2) imply
(3).  \epf

As a first order of business, we consider monic morphisms,
which capture the notion of a ``substructure''.
In categorical terms, a morphism $\varphi: \mathsf{C}_1\to \mathsf{C}_2$ is {\em monic} (or mono)
if for any other morphisms $\varphi_i : \mathsf{C}^* \to  \mathsf{C}_1$ ($i=1,2$) such that
$\varphi \circ \varphi_1 = \varphi \circ \varphi_2$, we have
$\varphi_1= \varphi_2$.

{\bf Remark}. To make a distinction in our reference to
morphisms at different levels, we reserve the term monic, mono, epi, etc for
Chu spaces, and use one-to-one, onto, injective, surjective for the functions
on the underlying sets. When properties on the underlying functions carry over to
Chu spaces, we occasionally mix the terms.

\begin{prop}\label{CB} We have:
\begin{enumerate}[\em(1)]
\item A morphism $\varphi: \mathsf{C}\to \mathsf{C}'$ in {\bf C} is monic  iff
$\varphi^+$ is injective and $\varphi^-$ is surjective.
\item A morphism $\varphi: \mathsf{C}\to \mathsf{C}'$ in {\bf E} is monic  iff
$\varphi^+$ is injective.
\item A morphism  $\varphi: \mathsf{C}\to \mathsf{C}'$ in ${\bf B}$ is monic  iff
$\varphi^+$ is injective.
\item Suppose $\varphi: (A, r, X)\to (B, s, Y)$ is a
morphism in  {\bf C} and $(B, s, Y)$ is extensional.
If $\varphi^+$ is surjective, then $\varphi^-$ is injective.
\end{enumerate}
\end{prop}

\pf
(1) The ``If'' part is straightforward. We check the ``Only If'' part. 
Suppose $\varphi : \mathsf{C}\to \mathsf{C}'$ is such that for any pair of
morphisms $\varphi_i : \mathsf{C}^* \to \mathsf{C}$, $i=1,2$, if $\varphi
\circ \varphi_1 = \varphi \circ \varphi_2$, then $\varphi_1 =
\varphi_2$.  We show that $\varphi^+$ is injective and $\varphi^-$ is
surjective.  Let's write $\mathsf{C}=(A, r, X)$ and $\mathsf{C}'=(A', r',
X')$. 

First we show that $\varphi^-$ is surjective. Suppose otherwise. 
Choose two sets $X_1, X_2$ of the same cardinality as $X \setminus
\varphi^-(X')$ such that $\varphi^-(X'), X_1, X_2$ are pairwise
disjoint.  Put $X^* = \varphi^-(X') \cup X_1 \cup X_2$ and $A^* = A$. 
For $i = 1,2$, choose a bijection $g_i: X \setminus \varphi^-(X') \to
X_i$ and let $g_i^* = id_{\varphi^-(X')}\cup g_i: X \to X^*$ be the
joint extension of the identity $id_{\varphi^-(X')}$ on $
\varphi^-(X')$ and $g_i$.  Let $f^* = id_A$.  For $a \in A^*$ and $x
\in X^*$, let $r^* (a, x) = r(a,x)$ if $x\in \varphi^-(X')$, $r^* (a,
x) = r(a, g_1^{-1}(x))$ if $x\in X_1$, and $r^* (a, x) = r(a,
g_2^{-1}(x))$ if $x\in X_2$. 

Let $\mathsf{C}^* = (A^*,r^*,X^*)$. Then $(f^*,g_i^*): \mathsf{C}^* \to {\sf
  C} $, with $i = 1,2$, are morphisms.  Indeed, $r(f^*(a), x) = r(a,
x) = r^*(a, g^*_i(x))$ if $x\in \varphi^-(X')$, and $r(f^*(a), x) = r(a,
x) = r^*(a, g_i(x))$ also, if $x\in X\setminus \varphi^-(X')$, for
$i=1,2$ by the definition of $r^*$.  Moreover, $(f^*,g_i^*)$ yield the
same composition with $\varphi$ because $g_i^*$s behave the same on
the image set $\varphi^-(X')$.  Now $\varphi$ being monic implies that
$g_1 = g_2$, a contradiction because $X \setminus \varphi^- (X') \ne
\emptyset$. Hence $\varphi^- (X') = X$.  Note that for this
construction to work, $\mathsf{C}^*$ cannot be required to be
extensional. 

The proof for the injectivity of $\varphi^+$ is the same as the ``only
if'' part for item (2), given next. 

(2) and (3).  \emph{If.} Let $\varphi : \mathsf{C}\to \mathsf{C}'$ be a morphism
and $\varphi^+$ an injection.  Consider two morphisms $\varphi_i :
\mathsf{C}^* \to \mathsf{C}$ ($i = 1,2$), which yield the same compositions
with $\varphi$.  Then $\varphi_1^+ = \varphi_2^+$, since $\varphi^+$
is injective.  By Prop.~\ref{lamarche}, we have $\varphi_1^- =
\varphi_2^-$.  Hence $\varphi$ is monic. 

\emph{Only If.} Let  $\varphi : \mathsf{C}\to \mathsf{C}'$  be monic and
write  $\mathsf{C} = (A,r,X )$ and  $\mathsf{C}' = (A',r',X')$. 
Assume  $a_1,a_2 \in A$  are such that  $\varphi^+ (a_1) = \varphi^+(a_2) = a'$. 
Construct a Chu space $\mathsf{C}^* = (A^*,r^*, X^*)$ as follows. 
Let  $A^* = \{a\}$, a singleton, and  $X^* = \Sigma$. 
Also, let  $r^*(a, \sigma )=\sigma$ for $\sigma \in \Sigma$. 
Clearly, $\mathsf{C}^*$  is biextensional. 

Now define $\varphi_i : \mathsf{C}^* \to \mathsf{C}$, $i = 1,2$, as follows. 
Let $\varphi^+_i(a) = a_i$ for $i = 1,2$.  For $x \in X$, put
$\varphi^-_i(x) = r(a_i, x) \in \Sigma = X^*$ $(i=1,2)$.  Then
$\varphi_1, \varphi_2 : \mathsf{C}^* \to \mathsf{C}$ are morphisms. 
Clearly, $\varphi^+ \circ \varphi^+_1 = \varphi^+ \circ \varphi^+_2$. 
We claim that also $\varphi^-_1 \circ \varphi^- = \varphi^-_2 \circ
\varphi^-$. Indeed, if $x' \in X'$ and $x = \varphi^-(x')$, then
$$
\varphi^-_1(x) = r(a_1,x) = r'(a', x')=r(a_2, x)= \varphi^-_2(x). 
$$

Hence  $\varphi_1, \varphi_2 : \mathsf{C}^* \to \mathsf{C}$
yield the same compositions with  $\varphi$. 
Since  $\varphi$  is monic, it follows that  $\varphi^+_1 = \varphi^+_2$. 
Thus  $a_1 = a_2$, and  $\varphi^+$  is injective. 

(4) Let $y,y' \in Y$ with $\varphi^-(y) = \varphi^-(y') = x$, say. Choose
any $b \in B$ and then $a \in A$ with $\varphi^+(a) =b$. Then
$s(b,y) = r(a,x) = s(b,y')$. Then $y=y'$ by extensionality.  \epf

The second and third items above would not be so surprising
  if the injectivity of $\varphi^+$ implied the surjectivity of
  $\varphi^-$ in {\bf B} and {\bf E}.  But this is not the case.

  \begin{ex} Consider $\mathsf{C}:= (\{ a \}, r, \{x_1, x_2\})$, with
  $r(a, x_1) =0$ and $r(a, x_2)=1$; $\mathsf{C}':= (\{ b \}, r', \{ y
  \})$, with $r'(b, y) =0$. Then the constraints $f(a) = b$ and $g(y)
  = x_1$ satisfy the property that $r'(f(a), y)= 0 = r(a,g(y))$ and
  the pair $(f,g)$ gives rise to a morphism. Clearly $\mathsf{C}, {\sf
  C'} \in {\bf B}$ and $f$ is injective, but $g$ is not surjective.
  With respect to items (2) and (3) in the proposition, this means
  that the backward component of a monic morphism in {\bf E} and {\bf
  B} need not be surjective.
\end{ex}

{\bf Remark}. Using a similar proof,
we can show that a morphism $\varphi: \mathsf{C}\to \mathsf{C}'$ is monic  if and only if
$\varphi^-$ is surjective, in the category of separable Chu spaces. We omitted
this statement in Prop.~\ref{CB} because we do not consider the category of
separable Chu spaces in the rest of the paper.

\section{Colimits of $\omega$-Chains}

\noindent We are interested in the subcategories of {\bf C}, {\bf E},
and {\bf B} with monic morphisms, denoted as {\bf iC}, {\bf iE}, and
{\bf iB}, respectively.  Let us begin with an unexpected observation
that colimits do not exist in {\bf iC} in general.  For this purpose,
we recall the definition of colimits, here formulated in {\bf iC}, but
it can easily be seen as an instantiation of a general
notion~\cite{maclane}.  We then show that colimits do exist in {\bf
  iE} and {\bf iB}.

\begin{defn}
An $\omega$-sequence in  ${\bf iC}$ is a family $(\mathsf{C}_i, \varphi_i)_{i\geq 1}$
$$\mathsf{C}_1 \cphi{\varphi}{1}  \mathsf{C}_2 \cphi{\varphi}{2} \mathsf{C}_3 \cphi{\varphi}{3}\cdots  \mathsf{C}_{i-1}\cphi{\varphi}{i-1} \mathsf{C}_i
\cphi{\varphi}{i}  \mathsf{C}_{i+1}\cdots $$
\end{defn}

\begin{defn}\label{cocone}
  A cocone from an $\omega$-sequence $(\mathsf{C}_i, \varphi_i)_{i\geq 1}$
  to a Chu space $\mathsf{C}:= (A, r, X)$ is a family of mappings ${\sf
    C}_i\cphi{\psi}{i} \mathsf{C}$ such that $\psi_{i+1} \circ \varphi_i
  = \psi_i,$ for all $i\geq 1$, i.e., the diagram

\[
\xymatrix @=7mm{
\mathsf{C}_1  \ar[r]^{{\varphi}_{1}}   \ar[rrdd] | {{\psi}_{1}}   &  \mathsf{C}_2
\ar[rdd] | {{\psi}_{2}}
  \ar[r]^{{\varphi}_{2}} &  \mathsf{C}_3 \ar[r]^{{\varphi}_{3}}
  \ar[dd] | {{\psi}_{3}}
  & \cdots
\ar[r]^>>>>{{\varphi}_{i}} &    \mathsf{C}_{i+1}
  \ar[lldd] | {{\psi}_{i+1}}
  \cdots
\\
  \\
& &    \mathsf{C}    & &
}\]
commutes.

A cocone $(\mathsf{C}_i\cphi{\psi}{i} \mathsf{C})_{i\geq 1}$ is universal if
for any other cocone $(\mathsf{C}_i\cphi{\psi'}{i} \mathsf{C}')_{i\geq 1}$
such that $ {\psi}^{'}_{i+1} \circ \varphi_i = {\psi}^{'}_i $ for all
$i\geq 1 $, there exists a unique $\mathsf{C}\cphi{\psi}{} \mathsf{C}'$ such
that $\psi\circ \psi_i = \psi^{'}_i$ for all $i\geq 1$.  Such a
universal cocone, if it exists, is called the colimit of the family $({\sf
  C}_i, \varphi_i)_{i\geq 1}$, while $\psi$ is called the mediating
map. In this case we write $\mathsf{C} = {{\rm
    co}\!\lim_{i}}(\mathsf{C}_i, \varphi_i)$.
\end{defn}

\begin{thm}\label{non-existence}
Colimits of $\omega$-chains of finite Chu spaces do not always exist in  {\bf iC}.
\end{thm}

\pf Consider finite Chu spaces $\mathsf{C}_i : = (\{1, \ldots , i\}, r_i, \{1,
\ldots , i\})$, such that $r_i(a,x) = 1$ if $a\leq x$, and $r_i(a,x) =
0$ otherwise. Observe that $\mathsf{C}_i$ is biextensional.  Define
$\varphi_i: \mathsf{C}_i \to \mathsf{C}_{i+1}$ such that $\varphi^+_i(a) =a$
for $a=1,\ldots, i$, and $\varphi^-_i(i+1) =i$, but $\varphi^-_{i}(x)
=x$ otherwise.   It is
straightforward to verify that the $\varphi_i$s are indeed morphisms: for
all $1\leq a\leq i$ and $1\leq x\leq i+1$,
$r_{i+1}(\varphi^+_i(a),x)=1$ iff $r_{i+1}(a,x)=1$ iff $a\leq x$ iff
$r_{i}(a, \varphi^-_i(x))=1$. Hence $\varphi_i$ is monic.

Consider $\mathsf{C}: = (\mathbb{N}, r, \mathbb{N})$, with $r(a,x)= 1$
iff $a\leq x$, and $r(a,x) = 0$ otherwise. Define $\psi_i : \mathsf{C}_i
\to \mathsf{C}$ by letting ${\psi_i}^+$ be inclusions and ${\psi_i}^- (x)
= x$ if $x\leq i$ and ${\psi_i}^- (x) = i$ for $x>i$.  One readily
checks that $(\psi_i : \mathsf{C}_i\to \mathsf{C})_{i \geq 1}$ is a cocone.
Consider another cocone defined by $\mathsf{C}': = (\mathbb{N}, r',
\mathbb{N}\cup\{t\} )$, where $r'$ extends $r$ with $r'(a,t)= 1$ for
all $a\in \mathbb{N}$. Define $\psi'_i : \mathsf{C}_i\to \mathsf{C}'$ by
${\psi'_i}^+ = {\psi_i}^+ $ and letting ${\psi'_i}^-$ extend
$\psi_i^-$ with ${\psi'_i}^- (t) = i$.  Clearly, $(\psi_i' : \mathsf{C}_i
\to \mathsf{C}')_{i \geq 1}$ is also a cocone.

Now we can infer that the colimit does not exist. More specifically,
suppose $(\psi_i^*: \mathsf{C}_i \to \mathsf{C}^*)_{i \geq 1}$ with ${\sf
  C}^* = (A^*, r^*, X^*)$ were a colimit.  First consider a mediating
map $\psi' : \mathsf{C}^* \to \mathsf{C}'$.  We have ${\psi'}^- (t) = x^*$
for some $x^*\in X^*$. Then for each $a^*\in A^*$ we obtain $r^*(a^*,
x^*) = r^*(a^*, \psi'^- (t)) = r'(\psi'^+(a^*), t) = 1$, thus $r^*(-,
x^*) =1$. Next consider a mediating map $\psi : \mathsf{C}^* \to {\sf
  C}$.  Since $\psi^-: \mathbb{N}\to X^*$ must be onto, $\psi^- (n) =
x^*$ for some $n\in \mathbb{N}$. We obtain
$0 = r(n+1, n) = r_{n+1}(n+1,n) = r_{n+1}(n+1, {\psi^*_{n+1}}^- (x^*))
= r^*({\psi^*_{n+1}}^+(n+1),x^*)= 1$, a contradiction.
\epf

Subsequently we will show that particular $\omega$-sequences of Chu
spaces do have colimits. For this we provide a generic
construction. It is the standard construction in the category of sets,
assimilated into the context of Chu spaces. We phrase it explicitly
since we will often refer to it.

\begin{const}\label{const:colimit}
  Let $(\mathsf{C}_i, \varphi_i)_{i \geq 1}$ be an $\omega$-sequence of
  Chu spaces where $\mathsf{C}_i = (A_i,r_i,X_i)$ and $\varphi_i^+:A_i
  \to A_{i+1}$ is the inclusion mapping, for each $i \geq 1$.
  Consider $\mathsf{C}:= (A, r, X)$ where
  \[\begin{array}{rcl}
    A & : = & \bigcup_{i\geq 1} A_i,\\
    X & : = & \{ (x_j)_{j\geq 1} \mid  \forall j\geq 1,\; x_j\in X_j \;{\&}\; \varphi^{-}_{j}(x_{j+1}) = x_j \},\\
    r(a,  (x_j)_{j\geq 1}) & : = & r_i(a,x_i) \text{ if } a \in A_i ~(i
    \geq 1).
  \end{array}\]

  \noindent Subsequently, we will denote a sequence $(x_j)_{j \geq 1} \in
  X$ often by $\tilde{x}$.

  For each $i\geq 1$, define $\mathsf{C}_i\cphi{\psi}{i} \mathsf{C}$ by
  $\psi^{+}_i (a) : = a $ and $\psi^{-}_i (\tilde{x}) : = x_i$ for all
  $a\in A_i$ and $\tilde{x}\in X$.
\end{const}

In Construction~\ref{const:colimit}, observe that possibly $X=
\emptyset$. Note that the relation $r$ is well-defined since if $i
\geq 1$ and $a \in A_i$, then $x_i = \varphi_i^-(x_{i+1})$ so
$r_{i+1}(a,x_{i+1}) = r_i(a,x_i)$; inductively we obtain $r_j(a,x_j) =
r_i(a,x_i)$ for each $j > i$.

Clearly, $\psi_i$ is a morphism.  Then we have, for each $\tilde{x}
\in X$,
  $$
  ( \varphi^{-}_i \circ \psi^{-}_{i+1}) (\tilde{x}) =
  \varphi^{-}_i(x_{i+1}) = x_i = \psi^{-}_i (\tilde{x})
  $$
  and for any $a\in A$,
  $$
  (\psi^{+}_{i+1}\circ \varphi^{+}_i) (a) = a = \psi^{+}_i(a)
  $$
  Therefore, $\psi_{i+1} \circ \varphi_i = \psi_i,$ and $({\sf
    C}_i\cphi{\psi}{i} \mathsf{C})_{i\geq 1}$ is indeed a cocone.  We note:

\begin{prop}
  If an $\omega$-sequence $(\mathsf{C}_i, \varphi_i)_{i \geq 1}$ in ${\bf
    iC}$ has a colimit, then this colimit is provided, up to
  isomorphism, by the cocone $(\mathsf{C}_i \cphi{\psi}{i} \mathsf{C})_{i \geq
    1}$ of Construction~\ref{const:colimit}.
\end{prop}

\pf
Let $(\mathsf{C}_i, \varphi_i)_{i \geq 1}$ have a colimit $(\mathsf{C}_i
\cphi{\psi'}{i} \mathsf{C}')_{i \geq 1}$ in ${\bf iC}$ where $\mathsf{C}' =
(A',r',X')$. By Proposition~\ref{CB}(1), the mappings $\varphi_i^+$
are injective and the mappings $\varphi^-_i$ are surjective, and we
may assume the $\varphi_i^+$s to be inclusions. Now construct ${\sf
   C}= (A,r,X)$ and $\psi_i: \mathsf{C}_i \to \mathsf{C} ~(i \geq 1)$. as in
Construction~\ref{const:colimit}. We claim that each $\psi_i$ $(i
\geq 1)$ is a morphism in ${\bf iC}$. By Proposition~\ref{CB}(1), it
remains to show that $\psi_i^-$ is onto. Using that the
$\varphi_j^-$s are onto, for any $x_i \in X_i$ we can easily find
$\tilde{x} \in X$ with $x_i = \psi_i^-(\tilde{x})$.

Since $\mathsf{C}'$ is the colimit, there is a unique $\psi:\mathsf{C}' \to
\mathsf{C}$ in ${\bf iC}$ such that $\psi \circ \psi_i' = \psi_i$ for
all $i \geq 1$. Then $\psi^+: A' \to A$ is injective. If $a \in A_i$
$(i \geq 1)$, then $a = \psi_i^+(a)=\psi^+ \circ {\psi_i'{}^+}(a)$ and
$\psi_i'{}^+(a) \in A'$, so $\psi^+$ is onto. Further, $\psi^-:X \to X'$
is onto, and we claim that $\psi^-$ is injective. Let $\tilde{x},
\tilde{y} \in X$ with $\psi^-(\tilde{x}) = \psi^-(\tilde{y})$.  For
each $i \geq 1$, then $x_i = \psi^-_i(\tilde{x}) = {\psi'_i}^- \circ
\psi^-(\tilde{x})= {\psi'_i}^- \circ \psi^-(\tilde{y}) =
\psi_i^-(\tilde{y}) =y_i$, showing $\tilde{x} = \tilde{y}$. Hence
$\psi$ is an isomorphism. \epf

In contrast to Theorem~\ref{non-existence}, we have the following.

\begin{thm}\label{limit-thm}
  Colimits exist in {\bf iE}, as given by Construction~\ref{const:colimit}.
\end{thm}


\pf Let $(\mathsf{C}_i, \varphi_i)_{i \geq 1}$ be an $\omega$-sequence in
{\bf iE} where $\mathsf{C}_i = (A_i,r_i,X_i)$ for each $i \geq 1$. By
Proposition~\ref{CB}(2), the mappings $\varphi_i^+$ are injective, and
we may assume the $\varphi_i^+$s to be inclusions. Now construct ${\sf
  C} = (A,r,X)$ and $\psi_i: \mathsf{C}_i \to \mathsf{C} ~(i \geq 1)$ as in
Construction~\ref{const:colimit}. We claim that $\mathsf{C}$ is
extensional. Let $\tilde{x}, \tilde{y} \in X$ and assume that
$r(-,\tilde{x}) = r(-, \tilde{y})$. We need to show that $\tilde{x} =
\tilde{y}$. Indeed, let $i \geq 1$ and choose any $a \in A_i$. Then
$r_i(a,x_i) = r(a, \tilde{x}) = r(a, \tilde{y}) = r_i(a,y_i)$.  So
$r_i(-,x_i) = r_i(-,y_i)$ and thus $x_i = y_i$ as $\mathsf{C}_i$ is
extensional. Hence $\tilde{x} = \tilde{y}$, and $\mathsf{C}$ is
extensional.



For universality, let $(\mathsf{C}_i\cphi{\psi'}{i} \mathsf{C}')_{i\geq 1}$
be a cocone, where $\mathsf{C}' = (A', r', X')$.  Define $\psi: \mathsf{C} \to
\mathsf{C}'$ by letting $\psi^+: A\to A'$ be such that $\psi^+(a) :=
{\psi'}^{+}_i(a)$ if $a\in A_i$, and letting $\psi^-: X'\to X$ be
given as $\psi^-(x') := ({\psi'}^{-}_m(x'))_{m\geq 1}$.  Then $\psi^+$
is well-defined because for any $1\leq i< j$, ${\psi '}^{+}_j(a) =
{\psi '}^{+}_i(a)$; also $\psi^{-}$ is well-defined because for any
$j\geq 1$, $\varphi^{-}_j ( {\psi '}^{-}_{j+1} (x')) = {\psi '}^
{-}_{j} (x')$, and hence the sequence $({\psi'}^{-}_m(x'))_{m\geq 1}$
belongs to $X$.  Further, $\psi$ is a morphism. We have $\psi \circ
\psi_i = {\psi '}_i$ for all $i\geq 1$ because $\psi^+(\psi^{+}_i(a))
= \psi^{+}(a) = {\psi '}_i(a)$, and $\psi^{-}_i(\psi^{-}(x') )=
\psi^{-}_i( ({ \psi '}^{-}_j(x'))_{j\geq 1}) = {\psi '}^{-}_i(x')$ for
all $a\in A_i$ and $x'\in X'$, by definitions.

The mediating morphism
$\psi$ is a morphism in {\bf iE} because   $\psi^+$ is injective,
and by Prop.~\ref{CB}(2), it is monic. The mediating morphism
is unique because its values are fixed by the commutativity requirements of the colimit diagram.
\epf

We now consider the biextensional case.
In order to avoid potential confusion of terminology, we call
a Chu space $\mathsf{C}:= (A, r, X)$ with finite $A$ and $X$ a finite Chu structure.
Finite objects in categorical terms will be studied in the next section, as we will
learn that finite objects and finite Chu structures do not always agree.

The proof of the following result involves a typical 
K{\"o}nig's-lemma argument for finite Chu structures which we will encounter
again later.

\begin{thm}\label{limit-thmB}
  Colimits exist in {\bf iB} for $\omega$-sequences of finite Chu structures. 
  They do not exist in general for $\omega$-sequences of arbitrary (non-finite) Chu structures in {\bf iB}.
   If for an
  $\omega$-sequence
  in ${\bf iB}$ there is a cocone to some Chu space in {\bf iB}, then
  the sequence has a colimit in {\bf iB}. 
\end{thm}

\pf The proof is similar to that of Theorem~\ref{limit-thm}, except
that we need to show that in case the $\mathsf{C}_i$ constitute a
sequence composed of finite biextensional structures, $\mathsf{C}:= (A,
r, X)$ is biextensional as well.  Since extensionality is already
shown to be preserved by this limit structure, it remains to show that
separability is preserved.

Suppose all $\mathsf{C}_i$s are separable.  Let $a,b \in A$ with $r(a,-)
= r(b,-)$. Let $m \geq 1$ be minimal with $a,b \in A_m$.  We claim
that $r_m(a,-) = r_m(b,-)$ in $\mathsf{C}_m$. The separability of ${\sf
  C}_m$ then gives us $a=b$. 

Suppose $r_m(a,-) \ne r_m(b,-)$. Then for any $i\geq m$ we have
$r_i(a,-) \ne r_i(b,-)$. For each element $x_i\in X_i$ such that
$r_i(a,x_i) \ne r_i(b,x_i)$ let $s(x_i) = (y_j)_{1 \leq j \leq i}$ be
the sequence defined inductively by $y_i=x_i$ and
$y_{j-1}=\varphi_{j-1}^-(y_j)$ for each $j =i , i-1, \ldots ,2$. 
Consider the set
$$T:=\{ s(x_i) \mid i \geq m \}.$$
Clearly, $T$ is an infinite set.  Consider elements of $T$ as nodes,
and an edge from $s (x_i)$ to $s (x_j)$ exists if $j=i+1$, and
$s(x_{i+1})$ is an extension of $s(x_i)$, that is, $x_i =
\varphi_i^-(x_{i+1})$. 
Observe that if $i \geq m$, $x_{i+1} \in X_{i+1}$, and $x_i =
\varphi_i^-(x_{i+1})$, then $r_i(a,x_i) \ne r_i(b,x_i)$ iff
$r_{i+1}(a,x_{i+1}) \ne r_{i+1}(b,x_{i+1})$. So, in this case $s(x_i)$
is defined iff $s(x_{i+1})$ is defined, and then they are connected by
an edge. Since each $X_i$ is finite, $T$ is a finite branching,
infinite tree.  By K\"{o}nig's Lemma, this tree has an infinite
branch, say, $(s(x_i))_{i \geq 1}$.  Consequently, $\varphi_i^-
(x_{i+1}) = x_i$ for all $i\geq 1$.  Clearly, by the construction
above, we have $\tilde{x}=(x_i)_{i\geq 1}\in X$ and $r(a,\tilde{x} )
\ne r(b, \tilde{x})$, a contradiction. 

For the second part of the theorem, we construct a counterexample as
follows. Let $\mathsf{C}_i := ( \mathbb{N}, r_i, \mathbb{N} )$ for $i\geq
1$, with $r_i (a, x) = 1$ iff $(x+i - 1) \mod a  = 0$ and
$r_i(a,x)=0$ otherwise. For example, $r_1$ is displayed below as
a countable matrix, which will be referred to as $\mathbb{M}$ for future
reference. 

\begin{center}
\begin{tabular}{| c c c  c  c  c c c c c  |}\hline
   1 & 1 & 1 & 1 & 1 & 1 & 1 &   \multicolumn{3}{c|}{$\cdots$} \\
   0 & 1 & 0 & 1 & 0 & 1 & 0 &   \multicolumn{3}{c|}{$\cdots$} \\
      0 & 0 & 1 & 0 & 0 & 1 & 0 &   \multicolumn{3}{c|}{$\cdots$}\\
            0 & 0 & 0 & 1 & 0 & 0 & 0 &   \multicolumn{3}{c|}{$\cdots$}\\
                        0 & 0 & 0 & 0 & 1 & 0 & 0 &   \multicolumn{3}{c|}{$\cdots$}\\
                         0 & 0 & 0 & 0 & 0 & 1 & 0 &   \multicolumn{3}{c|}{$\cdots$} \\
                               0 & 0 & 0 & 0 & 0 & 0 & 1 &  \multicolumn{3}{c|}{$\cdots$} \\
           \multicolumn{10}{|c|}{$\vdots$} \\ \hline
   \end{tabular}
\end{center}
Intuitively, $r_i$ is obtained by starting
from the countable matrix $r_1$ from the $i$-th column. Clearly, ${\sf
  C}_i$ is biextensional. The morphism $\varphi_i: \mathsf{C}_{i}\to {\sf
  C}_{i+1}$ is defined by $\varphi_i^+ := id_{ \mathbb{N}}$, and
$\varphi_i^- (x) = x+1$. Then,
$$r_{i+1} (\varphi_i^+(a), x) = r_{i+1}( a, x) = r_i(a, x+1) = r_i(a, \varphi_i^- (x)),$$
and so $\varphi_i$ is indeed a Chu morphism for each $i\geq 1$. 

Now assume that $(\psi_i:\mathsf{C}_i \to \mathsf{C})_{i \geq 1}$ is a cocone
where $\mathsf{C}=(A,r,X)$. We show that then $X = \emptyset$. Suppose
there is $x \in X$. Then, for any $i \geq
1$, $\psi_i^-(x)= \varphi_i^-(\psi_{i+1}^-(x))=\psi_{i+1}^-(x)+1$ by
the definition of $\varphi_i^-$. 
Inductively, we have $ \psi_{n+1}^-(x) = \psi_1^-(x) - n$ for all
$n\geq 1$.  But $ \psi_{n+1}^-(x) \in \mathbb{N}$ for all $n \geq 1$,
a contradiction.  Hence $X$ is empty. But then $\mathsf{C}$ is not
separable. 

For the last part of the theorem, let $(\mathsf{C}_i, \varphi_i)_{i \geq
  1}$ be an $\omega$-sequence in ${\bf iB}$ with some cocone $(\mathsf{C}_i
\cphi{\psi'}{i} \mathsf{C}')_{i \geq 1}$ in {\bf iB}. By
Theorem~\ref{limit-thm}, the sequence $(\mathsf{C}_i, \varphi_i)_{i \geq
  1}$ has a colimit $(\mathsf{C}_i \cphi{\psi}{i} \mathsf{C})_{i \geq 1}$ in
{\bf iE}. We claim that this is also the colimit of the sequence in
{\bf iB}. Now there is a morphism $\psi: \mathsf{C} \to \mathsf{C}'$ in {\bf
  iE} making the diagram commute. We show that $\mathsf{C}$ is separable. 
Let $\mathsf{C} = (A,r,X)$ and $\mathsf{C}' = (A',r',X')$. Choose any $a,b
\in A$ with $r(a,-)=r(b,-)$ . For any $x' \in X'$ we have
$r'(\psi^+(a),x') = r(a, \psi^-(x'))=r(b, \psi^-(x')) =
r'(\psi^+(b),x')$, so $\psi^+(a) = \psi^+(b)$ since $\mathsf{C}'$ is
biextensional, and $a=b$ as $\psi^+$ is injective. 

Hence $\mathsf{C}$ is biextensional, and by Proposition~\ref{CB}(2) and
(3), the morphisms $\psi_i: \mathsf{C}_i \to \mathsf{C}$ $(i \geq 1)$ belong
to {\bf iB}, and our claim follows. 
\epf

It is informative to think about the example given in the proof of
Theorem~\ref{non-existence}.  By the colimit construction, $\mathsf{C}'=
(\mathbb{N}, r', \mathbb{N}\cup\{t\} )$ is the colimit both in {\bf
  iE} and in {\bf iB}. 
There is indeed a monic morphism $\psi$ from $\mathsf{C}'$ to $\mathsf{C} =
(\mathbb{N}, r, \mathbb{N})$, where $\psi^+$ is the identity
(injection), and $\psi^-$ is the inclusion (but not onto). 

Also note that even though Theorem~\ref{limit-thm}
confirms that colimit always exists for extensional Chu spaces,
the counterexample for Theorem~\ref{limit-thmB} shows that
colimits for infinite structures may have weird behaviors with unintended
effects. This invites us to look more into objects constructed as colimits of finite structures,
in the next sections. 

\section{Finite objects}

\noindent In studying patterns of approximation in Chu spaces, finite
objects play an important role since they serve as the basis of
approximation.  In most cases, one expects finite objects to
correspond to finite structures, objects whose constituents are finite
sets.  In categorical terms, finite objects are captured using
colimits in a standard way, and the notion of ``approximation'' is
captured by monic morphisms.  Therefore, we work with categories {\bf
  iC}, {\bf iE}, and {\bf iB}.  However, since Prop.~\ref{CB}
indicates that what counts as monic morphisms depends on
extensionality, the existence of colimits and the characterization of
finite objects are not straightforward set-theoretic generalizations
obtained by treating each component of Chu spaces separately.

We give a characterization of the finite objects of ${\bf iC}$. 
Surprisingly, not all finite structures in ${\bf iC}$ are finite
objects; finite objects are characterized as extensional structures
with finite object set instead. 
The following definition is phrased in ${\bf iC}$; but as a general
categorical concept it can be made explicit in {\bf iE} and {\bf iB}
as well, and we do not repeat this here. 

\begin{defn}\label{fi-def}
  An object ~$\mathsf{F}$ of ~${\bf iC}$ is finite if for every
  $\omega$-sequence $(\mathsf{C}_i, \varphi_i)_{i\geq 1}$ of Chu spaces
  having a colimit, for every morphism $\varphi : \mathsf{F}\to {{\rm
      co}\!\lim_{i}}(\mathsf{C}_i, \varphi_i)$ in ${\bf iC}$
  there exist $i\geq 1$ and a morphism $\psi: \mathsf{F} \to \mathsf{C}_i$ such that the diagram
\[
\xymatrix @=10mm{ \mathsf{C}_1 \ar[r]^{\varphi_{1}} \ar[rrrdd] |
  {\psi_{1}} & \mathsf{C}_2 \ar[r]^{\varphi_{2}} \ar[rrdd] | {\psi_{2}} &
  \cdots \ar[r]^{\varphi_{i-1}} &\mathsf{C}_i \ar[r]^{\varphi_{i}}
  \ar[dd] | {\psi_{i}} & \cdots \ar[r]^<<<<<<{\varphi_{j}} & {\sf
    C}_{j+1} \ar[lldd] | {\psi_{j+1}} \cdots
  \\
  \\
  & & \mathsf{F} \ar[r]^<<<<<<<{\varphi} \ar@{-->}[ruu]^>>>>>>>>>>{\psi}
  & {{\rm co}\!\lim_{i}}(\mathsf{C}_i, \varphi_i) & & }\]
commutes, i.e., $\psi : \mathsf{F}\to \mathsf{C}_i$ is such that $\varphi =
\psi_i \circ \psi$. 
\end{defn}

If $\Sigma$ is finite and $\mathsf{F} = (A,r,X)$ is a finite object in
{\bf iC}, one can show that both $A$ and $X$ are finite sets,
i.e. $\mathsf{F}$ is a finite Chu space. However, somewhat surprisingly (at
least to us), the converse does not hold, as already simple examples
show, see Example~\ref{finite-example} below. The following result
characterizes the finite objects of {\bf iC}. 

\begin{thm}\label{both-finite}
  An object $\mathsf{F} = (B,s, Y)$ is finite in ${\bf iC}$ iff $B$ is
  finite and $\mathsf{F}$ is extensional. In this case, $|Y| \leq
  |\Sigma|^{|B|}$; in particular, if $\Sigma$ is finite, so is $Y$. 
\end{thm}


\pf \emph{(Only if.)}~  Suppose $\mathsf{F} = (B,s, Y)$ is a finite object of
${\bf iC}$ as in Definition~\ref{fi-def}.  If $B$ is infinite, then we
can write $B=A_1\cup \{a_i \mid i\geq 1\}$, where $A_1\cap \{a_i \mid
i\geq 1\} =\emptyset$.  Let $\mathsf{C}_i := (A_i, s_i , Y)$, where $A_i
= A_1\cup \{a_1, \ldots , a_i\}$ and $s_i$ is $s$ restricted to the
product $A_i\times Y$.  It can then be checked that $(\psi_i: {\sf
  C}_i \to \mathsf{F})_{i \geq 1}$ is the colimit of the
$\omega$-sequence $(\mathsf{C}_i, \varphi_i)_{i\geq 1}$, with
$\varphi^+_i, \psi^+_i$ inclusion and $\varphi_i^-, \psi_i^-$ identity
for all $i\geq 1$. We have $(id_B, id_Y): \mathsf{F}\to \mathsf{F}$ a
morphism.  Since $\mathsf{F}$ is a finite object, there exist an $i\geq 1$ and $\psi :
\mathsf{F}\to \mathsf{C}_i$, such that $id_B = \psi^+_i \circ \psi^+$. 
Then $a_{i+1} = \psi_i^+ \circ \psi^+(a_{i+1}) = \psi^+(a_{i+1}) \in
A_i$, a contradiction. 
Therefore, $B$ must be finite. 

Next we show that $\mathsf{F}$ is extensional.  Suppose there are
$y_1,y_2 \in Y$ with $y_1 \ne y_2$ and $s(-, y_1) = s(-, y_2)$. 
Put $Y'= Y \setminus \{y_1,y_2\}$. We may assume that $Y' \cap
\mathbb{N} = \emptyset$. Now let $\mathsf{C}_i = (B,s_i,Y_i)$ with $Y_i =
Y' \cup \{1, \ldots ,i\}$ and $\mathsf{C}=(B,s',Y' \cup \mathbb{N})$. Put
$s'(b,y) = s_i(b,y) = s(b,y)$ for each $y \in Y'$ and
$s'(b,j)=s_i(b,j)=s(b,y_1)$ for each $1 \leq j \leq i \in \mathbb{N}$
and $b \in B$. We let $\varphi_i^+$ be the identities, and
$\varphi^-_i$ leave everything unchanged except $\varphi^-_i (i+1) =
1$. 
Also, let $\psi^+_i$ be the identity map, $\psi^-_i (j) = j$ if $j
\leq i$ or $j\in Y'$, and $\psi^-_i (j) = 1$ if $j > i$. Then the
$\omega$-sequence $(\mathsf{C}_i, \varphi_i)_{i \geq 1}$ has $(\mathsf{C}_i
\cphi{\psi}{i} \mathsf{C})_{i \geq 1}$ as its colimit. Now define
$\varphi : \mathsf{F} \to \mathsf{C}$ with $\varphi^+ = id_{B}$, $\varphi^-$
mapping odd numbers to $y_1$, even numbers to $y_2$, and leaving
elements in $Y'$ unchanged.  Since $\mathsf{F}$ is finite, there are an
$i\geq 1$ and a morphism $\psi : \mathsf{F} \to \mathsf{C}_i$ such that
$\varphi = \psi_i \circ \psi$. 
Then $(\psi^- \circ \psi^-_i) (i+1) = \psi^- (1) = (\psi^- \circ
\psi^-_i) (i+2)$, contradicting the assumption that $\varphi^- (i+1)
\ne \varphi^- (i+2)$.

\emph{(If)}~ Suppose $\varphi: \mathsf{F} \to {{\rm
    co}\!\lim_{i}}(\mathsf{C}_i, \varphi_i)$ in {\bf iC}, where $B$ is
finite, $\mathsf{F}$ is extensional and $\mathsf{C}_i = (A_i,r_i,X_i)$ $(i
\geq 1)$. Since the colimit of an $\omega$-sequence is unique up to
isomorphism, we may assume that all $\varphi_i^+$s are inclusions and
that ${{\rm co}\!\lim_{i}}(\mathsf{C}_i, \varphi_i)$ is the
structure $(\mathsf{C}_i \cphi{\psi}{i} \mathsf{C})_{i \geq 1}$ with ${\sf
  C} = (A,r,X)$ given in Construction 3.2.  We can
further assume that $\varphi^+$ is an inclusion by renaming the
elements of $B$.  As $B$ is a finite set, $B \subseteq A=\bigcup_i
A_i$ implies that $B\subseteq A_m$ for some $m\geq 1$. 
Now define $\psi^-: X_m \rightarrow Y$ by $q\mapsto \varphi^-(x)$ if and only if
$x \in X$ satisfies $q=\psi^-_m(x)$. 

We show that  $\psi^-$ is a function. For this,
let $x,x'\in X$ such that
$\psi^-_m(x) = \psi^-_m(x')=q$.  Observe that $B \subseteq A_m
\subseteq A$.  For any $b \in B$, we have
\[\begin{array}{rclr}
   s (b,  \varphi^- (x))   &=& r (b,  x)~~~~~ & \mbox{($\varphi^+$ is inclusion)} \\
   &=& r_m (b,  \psi^-_m(x)) &  \mbox{($\psi^+_m$ is inclusion)} \\
   &=& r_m(b,  \psi^-_m(x'))~~~~~ & \mbox{($\psi^-_m(x) = \psi^-_m(x')$)} \\
   &=& r( b, x')  &  \mbox{($\psi^+_m$ is inclusion)} \\
   &=& s(b,  \varphi^- (x')) &  \mbox{($\varphi^+$ is inclusion)}\\
\end{array}\]
Therefore, $s(-, \varphi^-(x)) = s(-, \varphi^-(x'))$ and by extensionality,  $\varphi^-(x) = \varphi^-(x')$. 

\[
\xymatrix @=15mm{
  \mathsf{C}_m \ar[rd] | {{\psi}_{m}}   \ar[r] | {{\varphi}_{m}}  \ar@{-->}[d]^{{\psi^-}}     & \mathsf{C}_{m+1} \ar[d] | {{\psi}_{m+1}}     \\
  \mathsf{F} \ar@<0.8ex>[u]^{{\psi^+={\rm id}_B}} \ar[r] | {{\varphi}}
  & \mathsf{C}  \\
}\] (where $\mathsf{C}= {{\rm co}\!\lim_{i}}(\mathsf{C}_i,
\varphi_i$))

We check that  $\psi = ({\rm id}_B, \psi^-): \mathsf{F}\to \mathsf{C}_m$ is a Chu morphism. 
Indeed, for any $b\in B$ and $x_m \in X_m$, we have

\[\begin{array}{rclr}
r_m( \psi^+(b),  x_m)  &=& r_m( b,  x_m) &~~\mbox{($\psi^+={\rm id}_B$)}\\
&=& r_m (b,  \psi^-_m(x)) &~~\mbox{($\psi^-_m$  onto; $\psi^-_m(x)=x_m$)}\\
&=& r( \psi^+_m(b), x) &~~\mbox{($\psi_m: \mathsf{C}_m\to \mathsf{C}$)}\\
&=& r(b, x)  &~~\mbox{($\psi^+_m={\rm id}_{A_m}$)}\\
&=& r(\varphi^+(b), x)  &~~\mbox{($\varphi^+ = {\rm id}_{B}$)}\\
&=& s(b,  \varphi^-(x)) &~~\mbox{($\varphi: \mathsf{F}\to \mathsf{C}$)}\\
&=& s(b,  (\psi^-\circ \psi^-_m)(x)) &~~\mbox{($\psi^-\circ \psi^-_m =  \varphi^-$)}\\
&=& s(b,  \psi^-(x_m)) &~~\mbox{($\psi^-_m(x)=x_m$)}\\
\end{array}\]
as required. 
Since $\varphi^-$ is onto, $\psi^-$ is onto, and $\psi$ is
monic. Finally,
we have $\psi^- \circ \psi^-_m =  \varphi^-$ since for any $x\in X$,
by the definition of $\psi^-$, $\psi^-(\psi^-_m(x))=\varphi^-(x).$
Hence $\varphi = \psi_m \circ \psi$, and $\mathsf{F}$ is shown to be
finite. 

Finally, let $\Sigma^B$ denote the set of all functions from $B$ into
$\Sigma$. Note that $s(-,y) \in \Sigma^B$ for each $y \in Y$, and if
$\mathsf{F}$ is extensional, the mapping $y \mapsto s(-,y)$ provides an
injection of $Y$ into $\Sigma^B$, showing $|Y| \leq |\Sigma^B| =
|\Sigma|^{|B|}$ by cardinal arithmetic. \epf

Next we give two examples to illustrate Theorem~\ref{both-finite}. 

\begin{ex}\label{finite-example}
  Let $\Sigma = \{0,1\}$ and $\mathsf{F} : = (\{\star\},r, \{1,2\})$, a
  finite Chu space. If $r(\star,1) = r(\star,2) = 0$, then $F$ is not
  extensional and thus, by Theorem~\ref{both-finite}, not a finite
  object of ${\bf iC}$. To see this more explicitly, one can
  construct a sequence $(\mathsf{C}_i, \varphi_i)_{i \geq 1}$ as in the
  proof of Theorem~\ref{both-finite}, with $Y'= \emptyset$. 
\end{ex}

\begin{ex}\label{ex:Sig-infinite}
  Only in this example, let $\Sigma$ be an arbitrary (possibly
  infinite) set, and let $\mathsf{F} = (\{\star\}, r , \Sigma)$ with
  $r(\star,\sigma) = \sigma$ for each $\sigma \in \Sigma$. Then ${\sf
    F}$ is extensional, and by Theorem~\ref{both-finite}, $\mathsf{F}$ is a
  finite object of {\bf iC}. Trivially, if $\Sigma$ is infinite, {\sf
    F} is not a finite Chu space. 
\end{ex}


\begin{thm}\label{finite-E-onlyif}
  In the category ${\bf iE}$, if $\mathsf{F}= (B, s, Y)$ is finite then $B$
  is finite. 
\end{thm}

An independent proof is needed even though we follow a similar path as the proof of Theorem~\ref{both-finite}. 
Not only should we make sure that the Chu spaces involved are all extensional, but also the
monic morphisms are characterized differently. 
These entail non-trivial modifications from  the proof of Theorem~\ref{both-finite}. 

\pf Suppose $\mathsf{F}  = (B,s, Y)$ is a finite object in ${\bf iE}$. 
Suppose $B$ is infinite. Then we can write $B=A_1\cup \{a_i \mid
i\geq 1\}$, where $A_1\cap \{a_i \mid i\geq 1\} =\emptyset$. Fix $c
\in Y$. Let $\mathsf{C}_i := (A_1\cup \{a_1, \ldots , a_i\}, r_i ,
X_i)$, where $X_i = \{c\}$ and $r_i$ is $s$ restricted to the
product $(A_1\cup \{a_1, \ldots , a_i\})\times \{c\}$.  Clearly, all
$\mathsf{C}_i$s are extensional.  For morphisms $\varphi_i: \mathsf{C}_i
\to \mathsf{C}_{i+1}$, define $\varphi_i^+$ as inclusions, and
$\varphi_i^-: X_{i+1}\to X_i$ the identity.  By
Theorem~\ref{limit-thm}, the colimit $(\mathsf{C}_i \cphi{\psi}{i} {\sf
   C})_{i \geq 1}$ with $\mathsf{C} = (A, r, X)$ of the sequence $({\sf
   C}_i, \varphi_i)_{i \geq 1}$ exists, and can be taken as the one
given in Construction~\ref{const:colimit}.  Since each $X_i$ is a
singleton, $X$ is a singleton as well.  Thus we may assume $A = B$. 
With $\varphi^+$ identity and $\varphi^-$ inclusion, we obtain a
monic morphism $\varphi$ from $\mathsf{F}$ to $\mathsf{C}$.  Hence there is
a monic morphism $\psi$ from $\mathsf{F}$ to some $\mathsf{C}_i$ which
makes the required diagram commute. But then $\varphi^+(a_{i+1}) =
\psi_i^+(\psi^+(a_{i+1})) \ne a_{i+1}$, a contradiction.  \epf

The converse of Theorem~\ref{finite-E-onlyif} is not true.  To show
this, we adapt the counterexample for the second part of
Theorem~\ref{limit-thmB} as follows. Let $\mathsf{C}_i := ( \mathbb{N},
r_i, \mathbb{N}\cup \{c\} )$, with $r_i (a, x) = 1$ iff $(x+i - 1) \mod a = 0$ 
for $a,x \in \mathbb{N}$, and $r_i(-,c)=1$. 
Intuitively, $r_i$ is obtained by starting from the countable
matrix $r_1$ from the $i$-th column.  The morphism $\varphi_i: {\sf
  C}_{i}\to \mathsf{C}_{i+1}$ is defined by $\varphi_i^+ := id_{
  \mathbb{N}}$, and $\varphi_i^- (x) = x+1$, but we keep $c$ constant. 
Then, the colimit of this sequence is $\mathsf{C}:= ( \mathbb{N}, r,
\{c\} )$, with $r(-,c)=1$.  Now let ${\sf B}:= (\{1,2\}, s, \{c\})$,
with inclusion and identity paired to form a morphism $\varphi$ from
${\sf B}$ to $\mathsf{C}$.  There cannot be a morphism $\psi$ from ${\sf
  B}$ to any $\mathsf{C}_i$, because $\psi^-: (\mathbb{N}\cup \{c\}) \to
\{c\}$ cannot be defined, simply because ${\sf B}$'s column contains
two 1s, and each $r_i(2,i)=0$. Hence $B$ is a finite extensional Chu
space and thus a finite object of $\mathbf{iC}$ but not of
$\mathbf{iE}$. 

\begin{defn}
  A Chu space $(A,r,X)$ over $\Sigma$ is called \emph{discrete}, if
  for any mapping $f:A \to \Sigma$ there is $x \in X$ with $f
  =r(-,x)$. 
\end{defn}

\begin{thm}\label{finite-if}
  In the category ${\bf iE}$, $\mathsf{F}= (B, s, Y)$ is finite iff $B$ is
  finite and $\mathsf{F}$ is discrete. 
\end{thm}


\pf {\em (Only if)}.  By Theorem~\ref{finite-E-onlyif}, we know that
$B$ is finite.  Suppose $\mathsf{F}= (B, s, Y)$ is not discrete.  Let $v:
B\to \Sigma$ be such that $v\ne s(-, y)$ for any $y\in Y$.  Let ${\sf
  C}_i := (B\cup \mathbb{N}, r_i, Y\cup \mathbb{N} )$, with $r_i (a,
y) = s(a, y)$ for $a\in B$ and $y\in Y$, and $r_i (a, x) = 1$ iff 
$(x+i - 1) \mod a =  0$ for $a, x\in \mathbb{N}$. Furthermore, for all
$a\in \mathbb{N}$ and $y\in Y$, we let $r_i(a,y)=0$, and for all $b
\in B$ and $x\in \mathbb{N}$, we let $r_i(b,x)=v(b)$.  Viewed as an
infinite matrix, $r_i$ is obtained by placing $s$ at the upper-left
corner, and the infinite matrixes used in Theorem~\ref{limit-thmB} on
the lower-right corner. The lower-left corner is filled with zeros,
and the upper-right corner is filled with repeated columns duplicating
$v$.  A rendering of $r_1$ is given next, 
where $\mathbb{O}$ is a $\mathbb{N}\times Y$ matrix of all
$0$s, and $\mathbb{M}$ is the countable matrix used in the proof of
Theorem~\ref{limit-thmB}.

\begin{center}
\begin{tabular}{c c c  c  c  c c c}
& \multicolumn{3}{c}{$Y$} &  \multicolumn{4}{c}{$\mathbb{N}$} \\  \cline{2-8}
         \raisebox{-1.5ex}[0pt]{$B$}  &     \multicolumn{3}{|c|}{\raisebox{-1.5ex}[0pt]{$s$}} &
                           \multicolumn{1}{|c|}{\raisebox{-1.5ex}[0pt]{$v$}}
                              & \multicolumn{1}{|c|}{\raisebox{-1.5ex}[0pt]{$v$}} &
                             \multicolumn{1}{|c|}{\raisebox{-1.5ex}[0pt]{$v$}}
                                      & \multicolumn{1}{|c|}{\raisebox{-1.5ex}[0pt]{$\cdots$}}  \\
   &       \multicolumn{3}{|c|}{}   &  \multicolumn{1}{|c|}{}  &   \multicolumn{1}{|c|}{} &
                   \multicolumn{1}{|c|}{} &  \multicolumn{1}{|c|}{} \\   \cline{2-8}
       &       \multicolumn{3}{|c|}{}   &  \multicolumn{4}{|c|}{}  \\
\raisebox{0ex}[0pt]{$\mathbb{N}$}  &     \multicolumn{3}{|c|}{$\mathbb{O}$} &      \multicolumn{4}{c|}{$\mathbb{M}$}  \\
& \multicolumn{1}{|c}{}&\multicolumn{1}{c}{}&\multicolumn{1}{c|}{}&   \multicolumn{4}{c|}{}  \\   \cline{2-8}
   \end{tabular}
\end{center}

We define morphisms $\varphi_i: \mathsf{C}_i \to \mathsf{C}_{i+1} ~(i
\geq 1)$ by letting $\varphi_i^+ = id_{B \cup \mathbb{N}}$,
${\varphi_i^-}|_Y=id_Y$ and $\varphi_i^-(x) = x+1$ for each $x \in
\mathbb{N}$.  One can check that up to isomorphism the colimit of this
sequence is $(\psi_i: \mathsf{C}_i \to \mathsf{C})_{i \geq 1}$ where
$\mathsf{C}=(B \cup \mathbb{N}, s',Y)$ such that $s'$ coincides with
$s$ on $B \times Y$, and $s'(a,y) = 0$ for each $a \in \mathbb{N}$, $y
\in Y$; further, $\psi_i^+ = id_{B \cup \mathbb{N}}$ and $\psi_i^- =
id_Y$. Clearly, $\varphi=(id_B, id_Y)$ is a monomorphism from $\mathsf{F}$
to $\mathsf{C}$. Since $\mathsf{F}$ is finite, there are $i \geq 1$
and a morphism $\psi: \mathsf{F} \to \mathsf{C}_i$ which make the
diagram commute. Then $\psi^+=id_B$, and for any $b \in B$ and $x \in
\mathbb{N}$ we obtain $r_i(b,x)= s(b, \psi^-(x)) \ne v(b) = r_i(b,x)$,
a contradiction. 

{\em (If)}.  We follow a pattern  similar to the ``if'' part for
Theorem~\ref{both-finite}.  Suppose $\varphi: \mathsf{F} \to
{{\rm co}\!\lim_{i}}(\mathsf{C}_i, \varphi_i)$ in {\bf iE},
where $\mathsf{F} = (B, s, Y)$ is discrete and $B$ is finite. 
By Theorem~\ref{limit-thm},  we may
assume that ${{\rm co}\!\lim_{i}}(\mathsf{C}_i, \varphi_i)$
is the structure  $\mathsf{C}$ given in Construction~\ref{const:colimit}. 
  We can further assume that $\varphi^+$ is an
inclusion by renaming its elements.  As $B$ is a finite set, $B
\subseteq A=\bigcup_i A_i$ implies that $B\subseteq A_m$ for some
$m\geq 1$.  Since $\mathsf{F}$ is discrete, we can define $\psi^-: X_m
\rightarrow Y$ such that for each $x\in X_m$, $\psi^-(x) \in Y$ is
such that for all $b\in B$, $r_m(b, x) = s(b, \psi^-(x))$. 
This entails that $({\rm id}_B, \psi^-): \mathsf{F}\to \mathsf{C}_m$ is a
Chu morphism. 

To check the condition $\psi^- \circ \psi^-_m =  \varphi^-$, note
that for any $b\in B$ and $x \in X_m$, we have

\[\begin{array}{rclr}
s( b,  \psi^-(\psi_m^-(x) ) )  &=& r_m( b,  \psi_m^-(x) ) &~~\mbox{(def. of $\psi^-$)}\\
&=& r( \psi^+_m(b), x) &~~\mbox{($\psi_m: \mathsf{C}_m\to \mathsf{C}$)}\\
&=& r(b, x)  &~~\mbox{($\psi^+_m={\rm id}_{A_m}$)}\\
&=& r(\varphi^+(b), x)  &~~\mbox{($\varphi^+ = {\rm id}_{B}$)}\\
&=& s(b,  \varphi^-(x)) &~~\mbox{($\varphi: \mathsf{F}\to \mathsf{C}$)}\\
\end{array}\]
and by the extensionality of $\mathsf{F}$, we have $\psi^-(\psi_m^-(x) ) = \varphi^-(x)$, as required. 
  \epf

The following easy remark shows that the structure of discrete
extensional Chu spaces $\mathsf{C} = (A,r,X)$ is very restricted: it
is completely determined,  up to isomorphism, by the cardinality of the
object set $A$. 

\begin{rem}\label{rem:complext-iso}
  Let $\mathsf{C} = (A,r,X)$ and $\mathsf{C}'=(A',r',X')$ be two
  discrete extensional Chu spaces with $|A|=|A'|$. Then $\mathsf{C}$ and
  $\mathsf{C}'$ are isomorphic in {\bf iC}. 
\end{rem}

\pf
Choose a bijection $\varphi^+:A \to A'$. By the assumption on
$\mathsf{C}$, for each $x' \in X'$ there is a uniquely determined $x \in
X$ with $r(-,x) = r'(-,x') \circ \varphi^+$. The mapping $\varphi^-:
X' \to X$ with $\varphi^-(x') = x$ yields a Chu morphism
$\varphi=(\varphi^+, \varphi^-)$, and $\varphi^-$ is bijective by the
assumption on $\mathsf{C}'$. 
\epf

Almost similar to Theorem~\ref{finite-if}, we have the following. 
However, an independent proof is needed because
the structures used in the proof for Theorem~\ref{finite-if}
are not biextensional, and an extra case arises..

\begin{thm}\label{finite-if-finite}
   In the category ${\bf iB}$, $\mathsf{F}= (B, s, Y)$ is finite iff $B$
   is finite and $\mathsf{F}$ is discrete, or else $B$ is a singleton, $Y =
   \emptyset$, and $\Sigma$ is finite.
\end{thm}


\pf (\emph{If.}) ~In the first case, by Theorem 4.5, $\mathsf{F}$ is
finite in $\mathbf{iE}$, and $\mathsf{F}$ is biextensional. Note that
a colimit of an $\mathbf{iB}$-sequence taken in $\mathbf{iB}$
coincides with the colimit of this sequence taken in
$\mathbf{iE}$. Hence $\mathsf{F}$ is finite in $\mathbf{iB}$.

Secondly, assume $B$ is a singleton, $Y = \emptyset$, and $\Sigma$ is
finite.  Suppose $\varphi: \mathsf{F} \rightarrow {\rm
co}\!\lim_{i}(\mathsf{C}_i, \varphi_i)$ in $\mathbf{iB}$.  By Theorem
3.5, we may name $ {\rm co}\!\lim_{i}(\mathsf{C}_i, \varphi_i)$ as the
structure $\mathsf{C} = (A,r,X)$ given in Construction 3.2. Since
$\varphi^-: X \to Y$ is a mapping and $Y = \emptyset$, we also obtain
that $X = \emptyset$. Thus $A$ is a singleton, since $\mathsf{C}$ is
biextensional. Hence we may assume that $B = A = A_i$ for each $i \geq
1$.  Since each $\mathsf{C}_i$ is biextensional and $\Sigma$ is finite, we
obtain that $X_i$ is finite, too.

Suppose that $X_i \ne \emptyset$ for each $i \geq 1$.  For each $i
\geq 1$ and each element $x_i \in X_i$ define the sequence $s(x_i)$ as
in the proof of Theorem 3.5, and let again $T = \{s(x_i) | i \geq 1,
x_i \in X_i\}$. Then with the extension order, $T$ is an infinite
finite-branching tree since each $X_i$ is finite, and therefore $T$
contains, by K\"onig's Lemma, an infinite branch. This implies that $X
\ne \emptyset$, a contradiction.

Hence we have $X_i = \emptyset$ for some $i \geq 1$.  Then $\psi_i =
(\varphi^+,\emptyset): \mathsf{F} \to \mathsf{C}_i$ makes the diagram
commute, showing that $\mathsf{F}$ is finite.

(\emph{Only if.}) ~First we assume that $Y = \emptyset$.  Since ${\sf
F}$ is biextensional, $B$ must be a singleton, say, $B=\{b\}$. We
claim that $\Sigma$ is finite.  Suppose $\Sigma$ was infinite. We may
assume that $\mathbb{N} \subseteq \Sigma$.
We define Chu spaces  $\mathsf{C}_i = (A_i,r_i,X_i)$  such that
$A_i = B, X_i = \{j \in \mathbb{N}  \mid j \geq i\}$  and  $r_i(b,j) = 
j$  for each  $j \geq i$  and
$i \geq 1$. Also, let  $\varphi^+_i$  be the identity mapping and 
$\varphi^-_i$  be the inclusion.
Then  $(\mathsf{C}_i,\varphi_i)_{i \geq 1}$  is an $\omega$-chain in ${\bf 
iB}$  having  $\mathsf{F}$  as its
colimit. So there exists some  $i \geq 1$  such  that $\psi_i: \mathsf{F} 
\to \mathsf{C}_i$. In particular,
$\psi^-_i: X_i \to Y$  is a mapping, which contradicts the assumption 
that $Y = \emptyset$.

Next we assume that  $Y \ne \emptyset$  and we show that $\mathsf{F}$ is 
discrete.

 For this, we refine the arguments employed for
Theorem~\ref{finite-if}. If $\mathsf{F}=(B,s,Y)$ is not discrete,
choose $v: B \to \Sigma$ such that $v \ne s(-,y)$ for any $y \in Y$.
Now choose an infinite set $J$ of size at least $|B \cup Y|$.  Put $Y'
= \bigcup_{j \in J} Y \times \{j\}$; we write $Y_j$ for $Y \times
\{j\}$ for conciseness.

Let $\mathsf{C}_i := (B \cup J \cup \mathbb{N}, r_i, Y'
\cup\mathbb{N})$ $(i \geq 1)$ and $\mathsf{C}=(B \cup J \cup
\mathbb{N}, r, Y')$ where $r$ and $r_i$ $(i \geq 1)$ are defined as follows:

On each $B \times Y_j$ $(j \in J)$, define $r$ and $r_i$ precisely as
$s$ in $\mathsf{F}$ on $B \times Y$. Next, choose a bijection $\pi:J
\cup \mathbb{N} \to Y'$. On $(J \cup \mathbb{N}) \times Y'$ define $r$
and $r_i$ as ``unit matrix'', i.e.,  for any $x \in J \cup \mathbb{N}$
and $y \in Y'$ let $r(x,y) = r_i(x,y) = 1$ if $y = \pi(x)$, and
$r(x,y) = r_i(x,y)= 0$ otherwise. In the picture, $r_i$ is denoted as $\mathbb{I}_{\pi}$. 

Further, for all $b \in B$ and $x \in \mathbb{N}$ let $r_i(b,x) =
v(b)$. On $\mathbb{N} \times \mathbb{N}$, let $r_i$ be the same
relation as used in the proof of Theorem~\ref{limit-thmB}. Recall that
$(\mathbb{N}, r_i, \mathbb{N})$ is biextensional. Finally, put $r_i
\equiv 0$ on $J \times \mathbb{N}$.

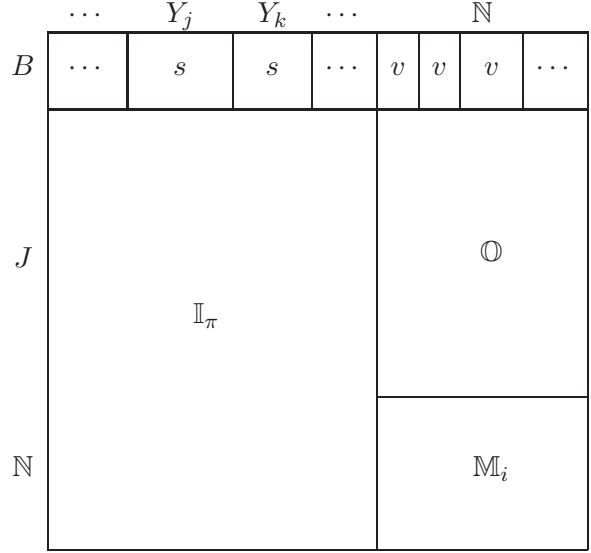
\begin{center}
\begin{tabular}{c ccc  ccc ccc cc  ccccc}
& \multicolumn{3}{c}{$\cdots$} &  \multicolumn{3}{c}{$Y_j$}  &  \multicolumn{3}{c}{$Y_k$}
                                       &  \multicolumn{2}{c}{$\cdots$}
                                       & \multicolumn{4}{c}{$\mathbb{N}$}
\\ \cline{2-16}
\raisebox{-1.5ex}[0pt]{$B$} &  \multicolumn{3}{|c|}{\raisebox{-1.5ex}[0pt]{$\cdots$}}
     & \multicolumn{3}{|c|}{\raisebox{-1.5ex}[0pt]{$s$}} & \multicolumn{3}{|c|}{\raisebox{-1.5ex}[0pt]{$s$}}
                                &  \multicolumn{2}{c}{\raisebox{-1.5ex}[0pt]{$\cdots$}}
                                      &   \multicolumn{1}{|c|}{\raisebox{-1.5ex}[0pt]{$v$}}
                                         &   \multicolumn{1}{|c|}{\raisebox{-1.5ex}[0pt]{$v$}}
                                            &   \multicolumn{1}{|c|}{\raisebox{-1.5ex}[0pt]{$v$}}
                                            & \multicolumn{1}{|c|}{\raisebox{-1.5ex}[0pt]{$\cdots$}} \\  
    & \multicolumn{1}{|c}{} &    \multicolumn{1}{c}{} & \multicolumn{1}{c|}{}   
    &  \multicolumn{1}{|c}{} &    \multicolumn{1}{c}{} & \multicolumn{1}{c|}{}     
    &  \multicolumn{1}{|c}{} &    \multicolumn{1}{c}{} & \multicolumn{1}{c|}{}     
       &  \multicolumn{1}{c}{}  &  \multicolumn{1}{c}{} 
       &                    \multicolumn{1}{|c|}{} & \multicolumn{1}{|c|}{}  & \multicolumn{1}{|c|}{} & \multicolumn{1}{c|}{} 
       \\  \cline{2-16}
     & \multicolumn{1}{|c}{} &    \multicolumn{1}{c}{} & \multicolumn{1}{c}{}      
    &  \multicolumn{1}{c}{} &    \multicolumn{1}{c}{} & \multicolumn{1}{c}{}      
    &  \multicolumn{1}{c}{} &    \multicolumn{1}{c}{} & \multicolumn{1}{c}{}     
       &  \multicolumn{1}{c}{}  &  \multicolumn{1}{c|}{} 
       &                    \multicolumn{1}{c}{} & \multicolumn{1}{c}{}  & \multicolumn{1}{c}{} & \multicolumn{1}{c|}{} 
       \\  
  \raisebox{-8ex}[0pt]{$J$}           & \multicolumn{1}{|c}{} &    \multicolumn{1}{c}{} & \multicolumn{1}{c}{}      
    &  \multicolumn{1}{c}{} &    \multicolumn{1}{c}{} & \multicolumn{1}{c}{\raisebox{-12.5ex}[0pt]{$\mathbb{I}_{\pi}$}}      
    &  \multicolumn{1}{c}{} &    \multicolumn{1}{c}{} & \multicolumn{1}{c}{}     
       &  \multicolumn{1}{c}{}  &  \multicolumn{1}{c|}{}     &  \multicolumn{1}{c}{}
       &  \multicolumn{1}{c}{}  & \multicolumn{1}{c}{\raisebox{-7.5ex}[0pt]{$\mathbb{O}$}}
        &   \multicolumn{1}{c|}{}     \\  
     & \multicolumn{1}{|c}{} &    \multicolumn{1}{c}{} & \multicolumn{1}{c}{}      
    &  \multicolumn{1}{c}{} &    \multicolumn{1}{c}{} & \multicolumn{1}{c}{}      
    &  \multicolumn{1}{c}{} &    \multicolumn{1}{c}{} & \multicolumn{1}{c}{}     
       &  \multicolumn{1}{c}{}  &  \multicolumn{1}{c|}{} 
       &                    \multicolumn{1}{c}{} & \multicolumn{1}{c}{}  & \multicolumn{1}{c}{} & \multicolumn{1}{c|}{} 
       \\  
            & \multicolumn{1}{|c}{} &    \multicolumn{1}{c}{} & \multicolumn{1}{c}{}      
    &  \multicolumn{1}{c}{} &    \multicolumn{1}{c}{} & \multicolumn{1}{c}{}      
    &  \multicolumn{1}{c}{} &    \multicolumn{1}{c}{} & \multicolumn{1}{c}{}     
       &  \multicolumn{1}{c}{}  &  \multicolumn{1}{c|}{} 
       &                    \multicolumn{1}{c}{} & \multicolumn{1}{c}{}  & \multicolumn{1}{c}{} & \multicolumn{1}{c|}{} 
       \\  \cline{13-16}
        \raisebox{-4.5ex}[0pt]{$\mathbb{N}$}           & \multicolumn{1}{|c}{} &    \multicolumn{1}{c}{} & \multicolumn{1}{c}{}      
    &  \multicolumn{1}{c}{} &    \multicolumn{1}{c}{} & \multicolumn{1}{c}{}      
    &  \multicolumn{1}{c}{} &    \multicolumn{1}{c}{} & \multicolumn{1}{c}{}     
       &  \multicolumn{1}{c}{}  &  \multicolumn{1}{c|}{}     &  \multicolumn{1}{c}{}
       &  \multicolumn{1}{c}{}  & \multicolumn{1}{c}{\raisebox{-4.5ex}[0pt]{$\mathbb{M}_i$}}
        &   \multicolumn{1}{c|}{}     \\  
     & \multicolumn{1}{|c}{} &    \multicolumn{1}{c}{} & \multicolumn{1}{c}{}      
    &  \multicolumn{1}{c}{} &    \multicolumn{1}{c}{} & \multicolumn{1}{c}{}      
    &  \multicolumn{1}{c}{} &    \multicolumn{1}{c}{} & \multicolumn{1}{c}{}     
       &  \multicolumn{1}{c}{}  &  \multicolumn{1}{c|}{} 
       &                    \multicolumn{1}{c}{} & \multicolumn{1}{c}{}  & \multicolumn{1}{c}{} & \multicolumn{1}{c|}{} 
       \\  
            & \multicolumn{1}{|c}{} &    \multicolumn{1}{c}{} & \multicolumn{1}{c}{}      
    &  \multicolumn{1}{c}{} &    \multicolumn{1}{c}{} & \multicolumn{1}{c}{}      
    &  \multicolumn{1}{c}{} &    \multicolumn{1}{c}{} & \multicolumn{1}{c}{}     
       &  \multicolumn{1}{c}{}  &  \multicolumn{1}{c|}{} 
       &                    \multicolumn{1}{c}{} & \multicolumn{1}{c}{}  & \multicolumn{1}{c}{} & \multicolumn{1}{c|}{} 
       \\  \cline{2-16}

    \end{tabular}
\end{center}
Observe that for each $b \in B$ either $s(b, -)$ is the constant-$0$
function, which implies that $r_i(b,-) = r(b,-) $ is constantly $0$ on
$Y'$, or else $r_i(b,y) = r(b,y)=1$ for infinitely many $y \in Y'$. It
follows that $\mathsf{C}_i$ and $\mathsf{C}$ are biextensional.

Now define morphisms $\varphi_i: \mathsf{C}_i \to \mathsf{C}_{i+1}$
and $\psi_i: \mathsf{C}_i \to \mathsf{C}$ such that $\varphi^+_i = \psi^+_i =
id_{B \cup J \cup \mathbb{N}}$, $\varphi_i^-$ and $\psi_i^-$ are the
identity on $Y'$, and $\varphi_i^-(x) = x+1$ for each $x \in
\mathbb{N}$. Then $(\psi_i: \mathsf{C}_i \to \mathsf{C})_{i \geq 1}$
is the colimit of the sequence $(\mathsf{C}_i, \varphi_i)_{i \geq 1}$. 
Next we define a morphism $\psi: \mathsf{F} \to \mathsf{C}$ by letting
$\psi^+ = id_{B}$ and $\psi^-(y,j)=y$ for each $y \in Y$, $j \in J$. 
Since $\mathsf{F}$ is finite, there is a morphism from $\mathsf{F}$
into some $\mathsf{C}_i$, which implies a contradiction by the choice
of $v$. 

Secondly, suppose that $B$ is infinite. For a subset $S \subseteq B$,
we call two elements $y,y' \in Y $ $S$-equivalent if $r(-,y)$ and
$r(-,y')$ coincide on $S$.  Now split $B = A' \cup \{ a_i \mid i \in
\mathbb{N} \} $ with $A' \cap \{a_i \mid i \in \mathbb{N}\} =
\emptyset$.  For each $i \geq 1$, let $A_i = A' \cup \{a_j \mid 1 \leq
j \leq i\}$, and let $X_i$ contain from each $A_i$-equivalence class
in $Y$ exactly one element. 

Let  $r_i$  be  $r$  restricted to  $A_i \times X_i$, and put
$\mathsf{C}_i = (A_i, r_i, X_i).$
Then  $\mathsf{C}_i$  is separable, since  $B$  is separable and  $X_i$  intersects
each  $A_i$-equivalence class, and  $\mathsf{C}_i$  is extensional since  $X_i$
contains from each  $A_i$-equivalence class at most one element. 

Now let  $\varphi_i^+$  be the identity, and for  $x_{i+1} \in X_{i+1} $
let  $\varphi_i^-(x_{i+1}) = x_i $ if  $x_i \in X_i $ lies in the
$A_i$-equivalence class of  $x_{i+1}$. 
The sequence  $(\mathsf{C}_i,\varphi_i)_{i \geq 1} $  has a colimit  $\mathsf{C} = (A,r,X)$
in  {\bf iE}, and clearly $\mathsf{C}$ is separable. Also, we may assume that  $\mathsf{C}$  is
obtained by Construction 3.4; then  $A = B$. 

There is a unique Chu morphism $\varphi$ from $\mathsf{F}$ to $\mathsf{C}$ with
$\varphi^+$ the identity on $B$; here the existence of $\varphi^-$
follows from $\mathsf{F}$ being discrete and the uniqueness from $\mathsf{F}$
being extensional.  Hence there are an $ i \geq 1$ and $\varphi: {\sf
F} \to \mathsf{C}_i$ which make the diagram commute, and this implies a
contradiction about $a_{i+1}$ as in the proof of
Theorem~\ref{both-finite}.  \epf

\section{Bifinite Chu spaces}

\noindent In this section, we will investigate Chu spaces which are,
intuitively and in a category-theoretic sense, countable objects and
approximable by the finite objects in the category. That is, we will
define bifinite Chu spaces as colimits of a sequence of (strongly)
finite Chu spaces. We will then show that this subcategory of
\textbf{iC} contains a universal homogeneous object.

Recall that the finite objects of \textbf{iC} may have an infinite
attribute set, if $\Sigma$ is infinite (cf. Example~\ref{ex:Sig-infinite}). 
For technical reasons (cf. the proofs
of Theorem~\ref{thm:colimit-icbif} and Proposition~\ref{amalg}),
we will need that the objects employed here have a finite and
non-empty set of attributes. We will call a space $\mathsf{F}$ in
\textbf{iC} \emph{strongly finite}, if $\mathsf{F}$ is a finite object
in \textbf{iC} and a finite Chu space with non-empty set of
attributes. Clearly, if $\Sigma$ is finite, the finite and the
strongly finite objects of $\mathbf{iC}$ with non-empty sets of
attributes coincide. 

\begin{defn}
  A Chu space in {\bf iC} is called bifinite if it is isomorphic to
  the colimit (with respect to {\bf iE}) of a chain of stronly finite
  objects in {\bf iC}.  The corresponding full subcategory of bifinite
  Chu spaces of {\bf C} and {\bf iC} are denoted as ${\bf C}_{\rm
    bif}$ and ${\bf iC}_{\rm bif}$, respectively. 
\end{defn}

As an example, consider the sequence of strongly finite biextensional
spaces $(\mathsf{C}_i,\varphi_i)_{i \geq 1}$ described in the proof of
Theorem~\ref{non-existence}.  As shown there, this sequence has no
colimit in the category $\mathbf{iC}$.  But by
Theorem~\ref{limit-thm}, the sequence has a colimit with respect to
the category $\mathbf{iE}$. This colimit thus belongs to
$\mathbf{iC}_{\rm bif}$.  Moreover, we will see below in
Theorem~\ref{thm:colimit-icbif}, that this space is also a colimit
of the given sequence with respect to the category $\mathbf{iC}_{\rm
  bif}$. 

Recall that any finite object of $\mathbf{iC}$ is extensional, hence
any bifinite Chu space is also extensional. It would not be
interesting to formulate the concept of bifinite spaces in \textbf{iE}
or \textbf{iB}, i.e. as colimits of chains of finite objects of
\textbf{iE} resp. \textbf{iB}: By Theorems~\ref{finite-if} and
\ref{finite-if-finite}, these finite objects are discrete. One can show that
colimits of chains of discrete extensional objects are again discrete
and extensional. Hence any two such `bifinite' objects (in \textbf{iE}
or \textbf{iB}) with countably infinite object set are isomorphic by
Remark~\ref{rem:complext-iso}. In contrast, we show that
$\mathbf{iC}_{\rm bif}$ is very large:

\begin{prop}
$\mathbf{iC}_{\rm bif}$ contains at least
continuum  many non-isomorphic objects. 
\end{prop}

\pf Consider a strictly increasing sequence of finite subsets $A_1
\subset A_2 \subset \ldots \subset \mathbb{N}$ of $\mathbb{N}$. We
define a sequence $(\mathsf{C}_i, \varphi_i)_{i \geq 1}$ as follows. 
For each $i \geq 1$, let $\mathsf{C}_i = ( A_i, r_i, X_i)$ with
$X_i=\{1, \ldots ,i\}$ and $r_i(a,j) = 1$ if $a \in A_j$ and $r_i(a,j)
= 0 $ otherwise, for any $a \in A_i$, $j \in X_i$. We let
$\varphi_i^+$ be the inclusion mapping, $\varphi_i^-(j) = j$ if $1
\leq j \leq i$, and $\varphi_i^-(i+1) = i$. As colimit of this
sequence of strongly finite objects we obtain, up to isomorphism,
$(\mathsf{C}_i \cphi{\psi}{i} \mathsf{C})_{i \geq 1}$ with
$\mathsf{C}=(\mathbb{N},r, \mathbb{N} \cup \{\infty\})$, $r(a,j)=1$ if
$a \in A_j$ and $r(a,j) = 0$ otherwise, for any $a,j \in \mathbb{N}$,
further $r(-, \infty)=1$, and $\psi_i^+$ inclusion, $\psi_i^-(j) = j$
if $1 \leq j \leq i$ and $\psi_i^-(j) = i$ if $i < j \in \mathbb{N}
\cup \{\infty\}$.  Note that in $\mathsf{C}$ the set $\{a \in
\mathbb{N} ~|~ r(a,i) = 1 \}$ equals $A_i$ if $i \in \mathbb{N}$, and
$\mathbb{N}$ if $i= \infty$. Hence the bifinite space $\mathsf{C}$
constructed in this way determines the sequence of subsets $(A_i)_{i
  \geq 1}$ uniquely, and two different sequences give rise to
non-isomorphic bifinite spaces.  Since there are continuum  many such
sequences, the result follows.  \epf

\begin{rem}
  By cardinality arguments, one can show that up to isomorphism
  $\mathbf{iC}_{\rm bif}$ has size $|\Sigma|^\omega$; this equals the
  continuum   if $\Sigma$ has size at most continuum. 
\end{rem}

With the restriction of objects to bifinite Chu spaces, colimits now
exist, in contrast to Theorem~\ref{non-existence}. 

\begin{thm}\label{thm:colimit-icbif}
  Colimits exist in ${\bf iC}_{\rm bif}$. 
\end{thm}

The technical content of the result is that $\mathbf{iC}_{\rm bif}$ is
closed in $\mathbf{iC}$ and in $\mathbf{iE}$ with respect to taking
colimits of sequences in $\mathbf{iC}_{\rm bif}$, and these colimits
taken in $\mathbf{iE}$ constitute the colimits of the given sequences
with respect to $\mathbf{iC}_{\rm bif}$. 

\pf 
First, let $(\mathsf{C}_i, \varphi_i)_{i \geq 1}$ be an
$\omega$-sequence in \textbf{iC} with finite objects
$\mathsf{C}_i=(A_i,r_i,X_i)$, $X_i \ne \emptyset$, and each
$\varphi_i^+$ being an inclusion. By Theorem~\ref{both-finite} each
$\mathsf{C}_i$ is extensional. Define $\mathsf{C}=(A,r,X)$ and
$\psi_i: \mathsf{C}_i \to \mathsf{C}$ $(i \geq 1)$ as in
Construction~\ref{const:colimit}. By Theorem~\ref{limit-thm}, the cocone
$(\mathsf{C}_i \cphi{\psi}{i} \mathsf{C})_{i \geq 1}$ is a colimit of
$(\mathsf{C_i},\varphi_i)_{i \geq 1}$ in $\mathbf{iE}$. We claim that
it is also a colimit of this sequence in $\mathbf{iC}_{\rm bif}$. 

First we show that each $\psi_i: \mathsf{C}_i \to \mathsf{C}$ is a
morphism in $\mathbf{iC}$, i.e., that $\psi_i^-: X \to X_i$ is onto. 
Choose any $x_i \in \mathsf{C}_i$. Clearly, since all $\varphi_j^-:
X_{j+1} \to X_j$ $(j \geq 1)$ are onto, there is a sequence $(x_j)_{j
  \geq 1}$ with $x_j \in X_j$ and $\varphi_j^-(x_{j+1})=x_j$ for each
$j \geq 1$. Then $\tilde{x} = (x_j)_{j \geq 1} \in X$ and
$\psi_i^-(\tilde{x}) =x_i$, as needed. 

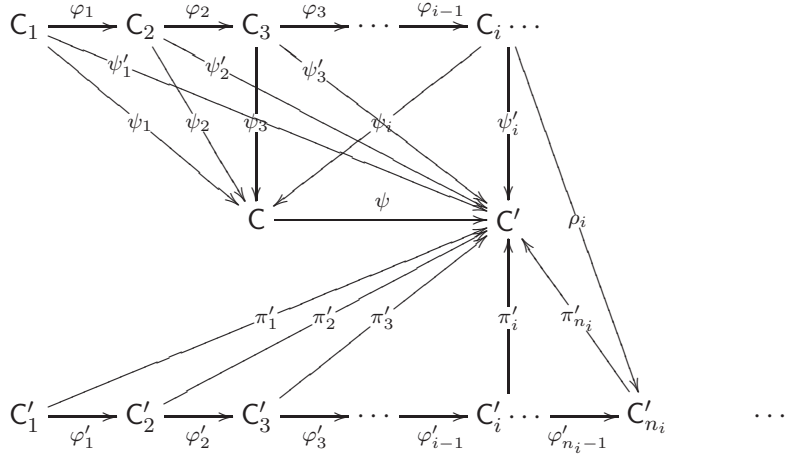
\[
\xymatrix @=9mm{ \mathsf{C}_1 \ar[r]^{{\varphi}_{1}} \ar[rrdd] |
  {{\psi}_{1}} \ar[rrrrdd] |<<<<<<<<<<{{\psi}'_{1}} & \mathsf{C}_2
  \ar[rdd] | {{\psi}_{2}} \ar[rrrdd] | <<<<<<<<{{\psi}'_{2}}
  \ar[r]^{{\varphi}_{2}} & \mathsf{C}_3 \ar[r]^{{\varphi}_{3}} \ar[dd] |
  {{\psi}_{3}} \ar[rrdd] | <<<<<<{{\psi}'_{3}} & \cdots
  \ar[r]^>>>>{{\varphi}_{i-1}} & \mathsf{C}_{i} \ar[lldd] |
  {{\psi}_{i}} \ar[dd] | {{\psi}'_{i}} \ar[rdddd] | {{\rho}_{i}}
  \cdots & &
  \\
  \\
  & & \mathsf{C} \ar[rr]^{{\psi}} & & \mathsf{C}' &&
  \\
  \\
  \mathsf{C}'_1 \ar[r]_{{\varphi}'_{1}} \ar[rrrruu] | {{\pi}'_{1}} & {\sf
    C}'_2 \ar[rrruu] | {{\pi}'_{2}} \ar[r]_{{\varphi}'_{2}} & {\sf
    C}'_3 \ar[r]_{{\varphi}'_{3}} \ar[rruu] | {{\pi}'_{3}} & \cdots
  \ar[r]_>>>>{{\varphi}'_{i-1}} & \mathsf{C}'_{i} \ar[uu] |
  {{\pi}'_{i}} \cdots \ar[r]_{\varphi'_{n_i-1}} & \mathsf{C}'_{n_{i}} \ar[luu] |
  {{\pi}'_{n_{i}}} & \cdots
  \\
}\]

For universality, let $\mathsf{C}'=(A',r',X') \in \mathbf{iC}_{\rm
  bif}$ and $(\mathsf{C}_i \cphi{\psi'}{i} \mathsf{C}')_{i \geq 1}$ be
a cocone.  Define $\psi: \mathsf{C} \to \mathsf{C}'$ as in the proof
of Theorem~\ref{limit-thm}. Then $\psi$ is a mediating morphism in
$\mathbf{iE}$, and it only remains to show surjectivity of $\psi^-: X'
\to X$. Choose any $\tilde{x} = (x_j)_{j \geq 1} \in X$.  Since ${\sf
  C}'\in \mathbf{iC}_{\rm bif}$, we can choose a chain of finite
objects $\mathsf{C}'_i = (A'_i,r'_i,X'_i)$ and monics $\varphi'_i$ such
that $(\mathsf{C}'_i \cphi{\pi'}{i}\mathsf{C}')_{i \geq 1}$ is a colimit of
this chain in ${\bf iC}$.  Again we can assume that the
${\varphi_i'}^+$s are inclusions and that the colimit $(\mathsf{C}'_i
\cphi{\pi'}{i} \mathsf{C}')_{i \geq 1}$ is given as in
Construction~\ref{const:colimit}.  Each $\mathsf{C}_i$ is a finite
object.  Observing the morphisms $\psi'_i: \mathsf{C}_i \to
\mathsf{C}' = {{\rm co}\!\lim_{}}_i(\mathsf{C}'_i, \varphi'_i)$, we
obtain a sequence of numbers $n_1 < n_2 < \ldots$ and monics $\rho_i:
\mathsf{C}_i \to \mathsf{C}'_{n_i}$ which make the diagram commute, that is,
for each $i \geq 1$ we have $\psi'_i = \pi'_{n_i} \circ \rho_{i}$.
For $i< j$ let $\varphi'_{i,j}$ be the composition of $\varphi'_i$ up
to $\varphi'_{j-1}$ from $\mathsf{C}'_i$ to $\mathsf{C}'_j$.  Since the
diagram commutes and the $\pi_i'$ are monic, we obtain
\[\varphi'_{n_i,n_{i+1}} \circ \rho_i   =  \rho_{i+1} \circ \varphi_i,  \mbox{~ for each ~} i \geq 1.\leqno{(*)} \]

Now consider the collection $K$ of all finite sequences $(x'_1,\ldots
,x'_m)$ such that
\[{x_j}' \in X'_{n_j}  \mbox{~and~}  \rho^-_j({x_j}') = x_j  \mbox{~ for each ~} j \leq m \leqno{(**)} 
 \]
and
$${\varphi'_{n_j, n_{j+1}}}^-(x'_{j+1}) = x'_j  \mbox{~for each~}  j < m.$$

We claim that there are arbitrarily long sequences. Choose any $m \in
\mathbb{N}$.  Since $\rho_j^-$ is surjective, there is $x'_m \in
X'_{n_m}$ satisfying $\rho^-_m(x'_m) = x_m$. Now put $x'_j =
{\varphi'_{n_j,n_m}}^-(x'_m)$ for each $1 \leq j < m$.  By (*),
inductively we obtain
$$
\rho^-_j(x'_j) = \rho_j^- \circ {\varphi'_{n_j,n_{j+1}}}^-(x'_{j+1}) =
\varphi^-_j \circ \rho^-_{j+1}(x'_{j+1}) = \varphi_j^-(x_{j+1}) = x_j,
$$
and our requirements (**) follow.

Now consider $K$ with the extension order. The sets $X'_j$ are finite
since the $\mathsf{C}'_j$ are strongly finite objects. 
So, $K$ is an 
infinite finite-branching tree. By K\"onig's lemma, there is
an infinite  branch in this tree. Hence there is an infinite sequence
$(x'_j)$ satisfying requirements (**) for each  $j \geq 1$. Now fill up this sequence
with the necessary additional elements from  $X'_i$  to obtain an element
$\tilde{x'} = (x''_j)_{j \geq 1} \in X'$ satisfying $x''_{n_j} = x'_j$
for each $j \geq 1$. We claim that $\psi^-(\tilde{x'}) =
\tilde{x}$. Indeed, choose any $i \geq 1$. Then
${\psi'_i}^-(\tilde{x'}) = \rho_i^- \circ {\pi'_{n_i}}^-(\tilde{x'}) =
\rho_i^-(x''_{n_i}) = \rho_i^-(x'_i) = x_i$. This proves our claim.

Secondly, consider an arbitrary $\omega$-sequence $(\mathsf{C}_i,
\varphi_i)_{i \geq 1}$ in $\mathbf{iC}_{\rm bif}$, and let
$(\mathsf{C}_i \cphi{\psi}{i} \mathsf{C})_{i^\geq 1}$ be its colimit
in $\mathbf{iE}$. We claim that this is also the colimit in
$\mathbf{iC}_{\rm bif}$. As before, we can show that each $\psi_i:
\mathsf{C}_i \to \mathsf{C}$ is a morphism in $\mathbf{iC}$. Now we
continue in a standard way. 

For each $i \geq 1$, since $\mathsf{C}_i \in \mathbf{iC}_{\rm bif}$,
we can write $\mathsf{C}_i$ as a colimit of a sequence of strongly
finite objects $(\mathsf{C}_{ij})_{j \geq 1}$ in $\mathbf{iC}$ and in
$\mathbf{iE}$. By a diagonal argument, there is a sequence of numbers
$n_1 < n_2 < \ldots$ such that the spaces $(\mathsf{C}_{i,n_i})_{i
  \geq 1}$ form a sequence in $\mathbf{iC}$ whose colimit, if it
exists, is also colimit of the sequence $(\mathsf{C}_i, \varphi_i)_{i
  \geq 1}$ in $\mathbf{iC}$, and whose colimit in $\mathbf{iE}$ is
$\mathsf{C}$. Hence $\mathsf{C} \in \mathbf{iC}_{\rm bif}$, and by our
first part, $\mathsf{C}$ is the colimit of the sequence
$(\mathsf{C}_{i,n_i})_{i \geq 1}$ also in $\mathbf{iC}_{\rm bif}$. Thus
$\mathsf{C}$ is the colimit of $(\mathsf{C}_i, \varphi_i)_{i \geq 1}$
in $\mathbf{iC}_{\rm bif}$.  \epf

To make the paper self-contained, we recall briefly a
result of Droste and G\"{o}bel~\cite{droste1} concerning the existence of a universal,
homogeneous object in an algebroidal category. 
Let ${\bf  G}$  be a  category  in which   all the morphisms   are  monic,
and ${\bf G}^*$ a full subcategory of ${\bf G}$.   Individually, an object $U$
of ${\bf G}$ is called
\begin{enumerate}[$\bullet$]
\item {\em ${\bf  G}^*$-universal} if
for any object $A$ in ${\bf G}^*$, there is a morphism $f:  A\to  U$;
\item  {\em ${\bf G}^*$-homogeneous}  if for any
$A$ in  ${\bf G}^*$ and any pair  $f, g: A\to U$,
there is an  isomorphism $h: U\to U$
such that $f=h\circ g$;
\end{enumerate}

Intuitively, $\mathbf{G}^*$-homogeneity means that each isomorphism
between two $\mathbf{G}^*$-substructures of $U$ extends to an
automorphism of $U$; this means that $U$ has maximal possible degree
of symmetry. 

Collectively, the category  $\mathbf{G}^*$ is said to have the
{\em amalgamation property} if for any $f_1:A\to B_1$, $f_2:A\to
B_2$ in ${\bf G}^*$, there exist $g_1:B_1\to B$, $g_2:B_2\to B$ in
${\bf G}^*$ such that $g_1\circ f_1 = g_2\circ f_2$. 

\begin{defn}
  Let $\mathbf{G}$ be a category in which all morphisms are monic. 
  Then $\mathbf{G}$ is called \emph{algebroidal}, if $\mathbf{G}$ has
  the following properties:

\begin{enumerate}[(1)]
\item $\mathbf{G}$ has a weakly initial object,
\item Every object of $\mathbf{G}$ is a colimit of an $\omega$-chain of
finite objects,
\item Every $\omega$-sequence of finite objects has a colimit, and
\item The number of finite objects of $\mathbf{G}$, up to isomorphism,
  is countable and between any pair of finite objects there exist only
  countably many morphisms. 
\end{enumerate}

\end{defn}

\begin{thm}\label{th:univ} (Droste and G\"{o}bel)
  Let ${\bf G}$ be an algebroidal category with all morphisms monic. 
  Let ${\bf G}_f$ be the full subcategory of finite objects of ${\bf
    G}$.  Then there exists a ${\bf G}$-universal, ${\bf
    G}_f$-homogeneous object iff $\mathbf{G}_f$ has the amalgamation
  property.  Moreover, in this case the ${\bf G}$-universal, ${\bf
    G}_f$-homogeneous object is unique up to isomorphism. 
\end{thm}

\begin{prop}\label{amalg} The category $\mathbf{iC}_{\rm bif}$
  contains an initial object. The strongly finite objects of
  $\mathbf{iC}$ are precisely the finite objects of $\mathbf{iC}_{\rm
    bif}$. If $\Sigma$ is countable, there are only countably many
  non-isomorphic finite objects in ${\bf iC}_{\rm bif}$. 
  Between any pair of finite objects there are only finitely
  many injections. Moreover, the finite objects of ${\bf
    iC}_{\rm bif}$ have the amalgamation property. 
\end{prop}

\pf The space $(\emptyset, \emptyset, \{x\})$ is the initial object of
$\mathbf{iC}_{\rm bif}$, since all spaces in $\mathbf{iC}_{\rm bif}$
have non-empty attribute sets. Next, we show only the amalgamation
property; the rest is easy to see.  Suppose $\mathsf{C}:=(A, r, X)$,
$\mathsf{C}_1:=(A_1, r_1, X_1)$, and $\mathsf{C}_2:=(A_2, r_2, X_2)$ are
strongly finite objects in {\bf iC} such that $A=A_1\cap A_2$, and let
$\varphi_1: \mathsf{C}\to \mathsf{C}_1$ and $\varphi_2: \mathsf{C}\to {\sf
  C}_2$ be morphisms in ${\bf iC}$. 

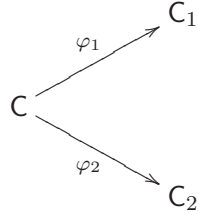
\[
\xymatrix @=7mm{
                        &      & \mathsf{C}_1    &        \\
\mathsf{C}     \ar[rru]^{\varphi_1}    \ar[rrd]_{\varphi_2}         &     &                   &        \\
                        &      & \mathsf{C}_2    &        \\
}\]
Construct $\mathsf{C}':=(A', r',  X')$ as:

\[\begin{array}{rcl}
A' & = & A_1\cup A_2\\
X' & = & \{(x_1, x_2) \in X_1 \times X_2 \mid \varphi^-_1(x_1) = \varphi^-_2(x_2)\} \\
r'( a,  (x_1, x_2)) & = & r_1( a, x_1)~~~~\mbox{\rm if $a\in A_1$}\\
r'( a,  (x_1, x_2)) & = & r_2( a, x_2)~~~~\mbox{\rm if $a \in A_2$}.\\
\end{array}\]
Note that in case $a\in A_1\cap A_2$,  we have
\[\begin{array}{rcl}
r_1(a, x_1 ) & = & r( a, \varphi^-_1(x_1) )\\
& =  & r (a, \varphi^-_2(x_2))\\
& = & r_2 (\varphi^+_2(a), x_2)\\
& =  & r_2( a, x_2).\\
\end{array}\]
To see that $\mathsf{C}'$ is extensional,
suppose $(x_1, x_2), (y_1, y_2) \in X' $ are such that
$r'(a, (x_1, x_2) ) = r'(a, (y_1, y_2))$ for all $a\in A_1\cup A_2$. 
By the definition of  $\mathsf{C}'$, then,
for each $a\in A_1$, we have
$r_1(a, x_1) = r_1(a, y_1)$. By the extensionality of  $\mathsf{C}_1$, we have
$x_1=y_1$. Similarly, by the extensionality of $\mathsf{C}_2$, we have
$x_2=y_2$ and so $ (x_1, x_2) = (y_1, y_2)$, as required. 
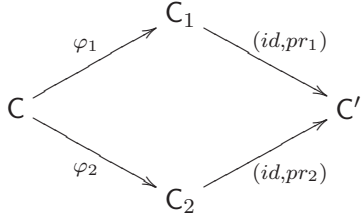
\[
\xymatrix @=7mm{
                        &      & \mathsf{C}_1   \ar[rrd]^{(id, pr_1)}  &  &      \\
\mathsf{C}     \ar[rru]^{\varphi_1}    \ar[rrd]_{\varphi_2}         &     &                   &   &  \mathsf{C}'    \\
                        &      & \mathsf{C}_2   \ar[rru]_{(id, pr_2)}   &    &    \\
}\]
It is easy to check further that
$(id, pr_1): \mathsf{C}_1\to \mathsf{C}'$
and  $(id, pr_2): \mathsf{C}_2\to \mathsf{C}'$
are morphisms in ${\bf iC}$. 
\epf

By Proposition~\ref{amalg}, the following result immediately follows from
Theorem~\ref{th:univ}. 

\begin{thm}\label{thm:iCbifalgebroidal}
  Let $\Sigma$ be countable. Then ${\bf iC}_{\rm bif}$ is an
  algebroidal category containing a universal homogeneous object $U$. 
  Moreover, $U$ is unique up to isomorphism. 
\end{thm}

  Since ${\bf iC}_{\rm bif}$ contains spaces with an attribute set of
  size continuum, it follows that the attribute set of $U$ also has
  size continuum. However, we just note that since the proof of
  Theorem~\ref{th:univ} is constructive, we can
  \emph{construct a sequence} $(\mathsf{C}_i,\varphi_i)_{i \geq 1}$ whose
  colimit is the universal homogeneous object $U$.

We remark that ${\bf iC}_{\rm bif}$ does not contain all countable extensional Chu spaces
(just as not all countable cpos are SFP). 
Let  $\mathsf{C} = (\mathbb{N},r,\mathbb{N})$
be the biextensional Chu space
described in the proof of Theorem 3.1. We claim that
$\mathsf{C}$  is not bifinite. 

Indeed, choose the sequence $(\mathsf{C}_i,\varphi_i)_{i \geq 1}$, the
space $\mathsf{C}' = (\mathbb{N},r',\mathbb{N} \cup \{t\})$ and the
monics $\psi'_i: \mathsf{C}_i \to \mathsf{C}'$ as in the proof of Theorem
3.1.  By Theorem 3.4, $(\psi'_i:\mathsf{C}_i \to \mathsf{C}')_{i \geq
  1}$ is the colimit of the chain $(\mathsf{C}_i,\varphi_i)_{i \geq 1}$
in ${\bf iE}$.  By Theorem 5.3, this is also the colimit of the
sequence of finite spaces $\mathsf{C}_i$ in the category ${\bf iC}_{\rm
  bif}$.  Consider the morphisms $(\psi_i: \mathsf{C}_i \to {\sf
  C})_{i\geq 1}$ described in the proof of Theorem 3.1.  Now if ${\sf
  C}$ was bifinite, there would be a unique morphism $\psi: \mathsf{C}'
\to \mathsf{C}$ in ${\bf iC}_{\rm bif}$ making the diagram commute. But
then $\psi^+ = id_{\mathbb{N}}$, and $\psi^-(n) = t$ for some $n \in
\mathbb{N}$, yielding $0 = r(n+1,n) = r'(n+1,t) = 1$, a contradiction. 

\section{Concluding Remarks}

\noindent One specific motivation for considering the notion of
bifinite Chu space is for modeling linear logic~\cite{lamarche0}, for
which tensor and linear negation should also be brought into the
picture.  Although the category of bifinite Chu spaces is monoidal, it
is not monoidal closed~\cite{huang}.  Also, the construction of linear
negation cannot be accounted for nicely in bifinite Chu spaces either.
In spite of these, one has to look at the bigger picture and consider
our results in the context of Chu spaces as a general framework for
studying the dualities of objects and properties, points and open
sets, and terms and types, under rich mathematical contexts. This view
has already been made amply clear by
Pratt~\cite{pratt1,pratt2,pratt3,pratt4,pratt5,pratt6,pratt7}.
Traditionally, the study on Chu spaces had a non-constructive
flavor. This paper provides (1) a basis for a more constructive
analysis of categories of Chu spaces collectively; (2) a framework in
which finite Chu spaces can be used to approximate infinite ones as
colimits of $\omega$-chains of finite Chu spaces; (3) the
``completeness'' of the delineated categories under the colimit
construction.  On the technical side, the development of our framework
hinges upon the adoption of monic morphisms as the basic steps for
approximation at the structural level.  It is certainly reasonable to
consider other possible notions of ``approximation'' as well, such as
``regular mono'' or more generally ``embedding-projection pair'', of
which monic morphism can be regarded as a special case.

\bigskip
{\bf Acknowledgment}. The authors would like to thank F. Lamarche and the anonymous 
referees for valuable feedback.

\end{document}